\documentclass[10pt]{article}
\usepackage[left=2.5cm, right=2.5cm, top=2.5cm, bottom=2.5cm]{geometry}
\usepackage{setspace}
\usepackage{booktabs}
\usepackage{siunitx}
\usepackage[colorlinks,linkcolor=black,anchorcolor=black,citecolor=blue]{hyperref}
\usepackage{amsmath, amsfonts, amssymb} 
\usepackage{multirow}  
\usepackage{array}      
\usepackage{tablefootnote}
\usepackage[version=3]{mhchem} 
\usepackage{lastpage,fancyhdr,graphicx}
\usepackage{epstopdf}
\usepackage{color}
\usepackage{sidecap}
\usepackage{wrapfig}
\usepackage[pscoord]{eso-pic}
\usepackage[fulladjust]{marginnote}
\usepackage{amsmath}  
\usepackage{threeparttable} 
\usepackage{makecell}
\usepackage[labelsep=period]{caption}
\usepackage{caption}
\usepackage{float}
\reversemarginpar

\usepackage[backend=biber,style=nature,sorting=none]{biblatex}
\begin{document}

\begin{flushleft}
{\huge\bfseries Heralded Emission Detection in InAs/ZnSe Quantum Dot Solids Using Time-Correlated Photons}

\vspace{0.3cm}

{\large
Chieh Tsao\textsuperscript{1,2}, 
Xiang Li\textsuperscript{3,4},
Alex Hinkle\textsuperscript{1},
Yifan Chen\textsuperscript{1}, 
Elvar Oskarsson\textsuperscript{1}, 
Uri Banin\textsuperscript{3,4}, 
Hendrik Utzat\textsuperscript{1,2*}
}
\\
\bigskip
\textsuperscript{\textit{1}} Department of Chemistry, University of California, Berkeley, California 94720, USA
\\
\textsuperscript{\textit{2}} Materials Science Division, Lawrence Berkeley National Lab, Berkeley, California 94720, USA
\\
\textsuperscript{\textit{3}} Institute of Chemistry, The Hebrew University of Jerusalem, Jerusalem 91904, Israel
\\
\textsuperscript{\textit{4}} The Center for Nanoscience and Nanotechnology, The Hebrew University of Jerusalem, Jerusalem 91904, Israel
\\
* \href{mailto:hutzat@berkeley.edu}{hutzat@berkeley.edu}

\end{flushleft}


\section*{Abstract} 
Harnessing quantum correlations between photons is an emerging frontier in optical spectroscopy, yet experimental demonstrations have largely remained limited to molecular systems at room temperature. Here, we investigate heralded emission detection (HED) under continuous-wave entangled-photon excitation of near-infrared (NIR)–emitting colloidal III–V quantum dots (QDs) solids at low temperatures. We demonstrate the advantages of superconducting nanowire single-photon detectors (SNSPDs) for high time resolution ($\sim$72 ps) and large-area NIR avalanche photodiodes (APDs) for high emission count rates ($\sim$2000 cps). Second-order photon-correlation analysis reveals exciton lifetimes and fine-structure energy splittings. These results establish NIR colloidal QDs as a bright, tunable model system for quantum-light spectroscopy and highlight their compatibility with optical cavities as a further experimental control parameter.



\begin{refsection}
\section{Introduction}
Optical spectroscopy continues to illuminate foundational processes in chemical, biomolecular, and solid-state systems across timescales. Thus far, canonical optical spectroscopy uses classical light from pulsed or continuous-wave (CW) lasers producing coherent wavepackets or wavetrains. The conventional semiclassical description for the light-matter interaction, treating light as a classical perturbation of a quantum system, is therefore typically sufficient \cite{scully1997quantum, boyd2008nonlinear}. Electronic or vibrational coherences generated through pulse/sample interaction can be interfered with additional time-delayed optical pulses, thereby bestowing non-linear spectroscopy with exquisite control via the bandwidth, shape, and delay between pulses \cite{weinacht2018time}. These control knobs can extract the energy-time correlations of a sample response without requiring any quantization of field excitations, i.e., without invoking photons.

The use of quantum light, i.e. photons with defined correlations in time, energy, or polarization \cite{dorfman2016nonlinear,mukamel2020roadmap} is an emerging new paradigm in optical spectroscopy. This is distinctly different from classical spectroscopy, where coherent laser light is characterized by a Poissonian number distribution of otherwise uncorrelated photons \cite{fox2006quantum, gerry2023introductory}. Quantum spectroscopy can therefore use what classical spectroscopy cannot; the inherent correlations between photons in bespoke quantum states. Recent theoretical efforts have explored non-classical correlations between photons in spectroscopic experiments e.g., using energy-time entangled pairs. Intriguing theoretical predictions \cite{dorfman2016nonlinear} beyond overcoming classical noise limits have been made \cite{aasi2013enhanced}. For instance, access to classically forbidden excitation pathways \cite{muthukrishnan2004inducing}, control over photo-chemical conversion rates through vibrational wavepacket preparation \cite{gu2021photoisomerization, zhou2025enhanced}, or remote-controlled spectroscopy via twin-pair detection of entangled states have been proposed \cite{fujihashi2023probing}.

Only a few experimental studies on entangled-photon light-matter interaction follow the theoretical lead. As such, the optical response of materials and molecules under quantum light excitation, e.g., single-photon Fock states or energy-time entangled photon pairs from spontaneous parametric downconversion (SPDC) has gradually come into focus \cite{couteau2018spontaneous, anwar2021entangled}, but is still limited by the significant experimental complexity of quantum light spectroscopy. In this context, the still-limited brightness of entangled photon sources used for sample excitation is a particular challenge \cite{tsao2025enhancing}. This is different from most spectroscopic experiments, in which sample stability rather than the achievable excitation power is the limiting factor. Probing sample coherences after entangled-photon interaction with the sample using a 'probe' pulse (or second photon) is therefore not easily achieved due to the low probability of two photons interacting with the same emitter. Instead, either the entangled photon phase delay (in transmission) or spontaneous emission as proxy for entangled photon absorption is measured.

Two common entangled two-photon (ETP) spectroscopies include fluorescence-detected two-photon absorption \cite{burdick2021enhancing,varnavski2020two, gu2020manipulating} and heralded emission detection (HED) \cite{harper2023entangled,eshun2023fluorescence,li2023single,alvarez2025correlated,gabler2025benchmarking}. Two-photon absorption using ETP has been reported to exhibit stronger absorption compared to classical two-photon absorption, owing to the intrinsic time correlation of photons in pairs \cite{burdick2021enhancing}. This effect has potential applications in fluorescence microscopy \cite{varnavski2020two}, although it remains somewhat debated \cite{hickam2022single}. HED leverages the intrinsic time correlation of photon pairs from quantum light sources such as SPDC. One photon excites the emissive sample, while its twin is directly detected to “herald” the excitation event. Correlating the herald with the following photoluminescence photon reconstructs the emission dynamics, using the twin-pair correlations. This is fundamentally different from  the laser-pulse correlations employed in conventional lifetime measurements using time-correlated single-photon counting (TCSPC). HED lifetime measurements therefore work with pulsed and CW lasers. The wavelengths of the entangled photons can be finely tuned by adjusting the temperature of the nonlinear crystal in the SPDC process, allowing for wavelength-dependent excitation studies. Because SPDC produces photon statistics in each arm that follow a thermal distribution, HED can be employed to study emission lifetimes under thermal light-like conditions. This makes it especially valuable for exploring systems such as photosynthetic complexes or photovoltaic materials under illumination with realistic underlying photon statistics. Another key difference is that TCSPC requires very low detection probability per pulse (1–5\%) to avoid pile-up, which limits usable counts \cite{Lakowicz2006}. HED, not tied to an excitation clock, avoids pile-up and can in principle operate at higher fluxes. Despite its conceptual simplicity, only a few studies have applied HED, all of which have measured solution-phase molecular systems such as Indocyanine Green (ICG) \cite{harper2023entangled}, Rhodamine 6G (R6G) \cite{eshun2023fluorescence}, and IR-140 \cite{gabler2025benchmarking}. A noteworthy feature of HED is its ability to herald single-photon Fock states to study light-driven processes in the ultimate limit of excitation intensity, e.g., in Light-Harvesting 2 (LH2) complexes \cite{li2023single, alvarez2025correlated}.

To advance quantum spectroscopy, both the scope of accessible methods and samples must be expanded. HED's current main limitation, shared with all other entangled-pair spectroscopy, is the low photon flux and coincidence rates from entangled-photon sources. HED is therefore currently not competitive for routine lifetime measurements. However, it can provide a foundation for extending quantum-light spectroscopy beyond simple intensity correlations, opening the possibility to resolve additional energy- and polarization-correlations. Such capabilities would enable more sophisticated approaches for probing quantum systems, with the broader aim of accessing entanglement signatures and unraveling excitonic dynamics in quantum materials \cite{saglamyurek2011broadband}. In light of these requirements, the expansion of entangled-pair spectroscopy from solution-phase organic molecules to solid-state materials is a logical next step. First, the entangled light-matter interaction and emission can be photonically enhanced to increase both the overall brightness and excitation rate per emitter. The latter is needed, for example, to validate theoretical predictions of inter-pair delay-dependent two-photon excitation i.e., electronic pathway selectivity \cite{fujihashi2024pathway}.

Here, we demonstrate HED in solid-state materials at low temperatures, specifically thin-films of near-infrared (NIR) emitting colloidal InAs/ZnSe quantum dots (QDs). QDs offer a rich platform for exploring entangled-pair light–matter interactions, thanks to their tunable electronic fine structure, phonon coupling, and electron–hole radiative and non-radiative rates \cite{bera2010quantum}. To our knowledge, our results represent the first demonstration of entangled-photon heralded emission detection in a solid-state material, distinct from prior two-photon excitation approaches using entangled photons \cite{pandya2024towards}. QD's processability and proven integration with nanophotonic structures may help overcome brightness limitations through enhancement of the absorption and emission \cite{chen2021integration}, and serve as a model system for entangled-photon non-linear spectroscopy.

\section{Result}
\begin{figure}
    \centering\includegraphics[width=16cm]{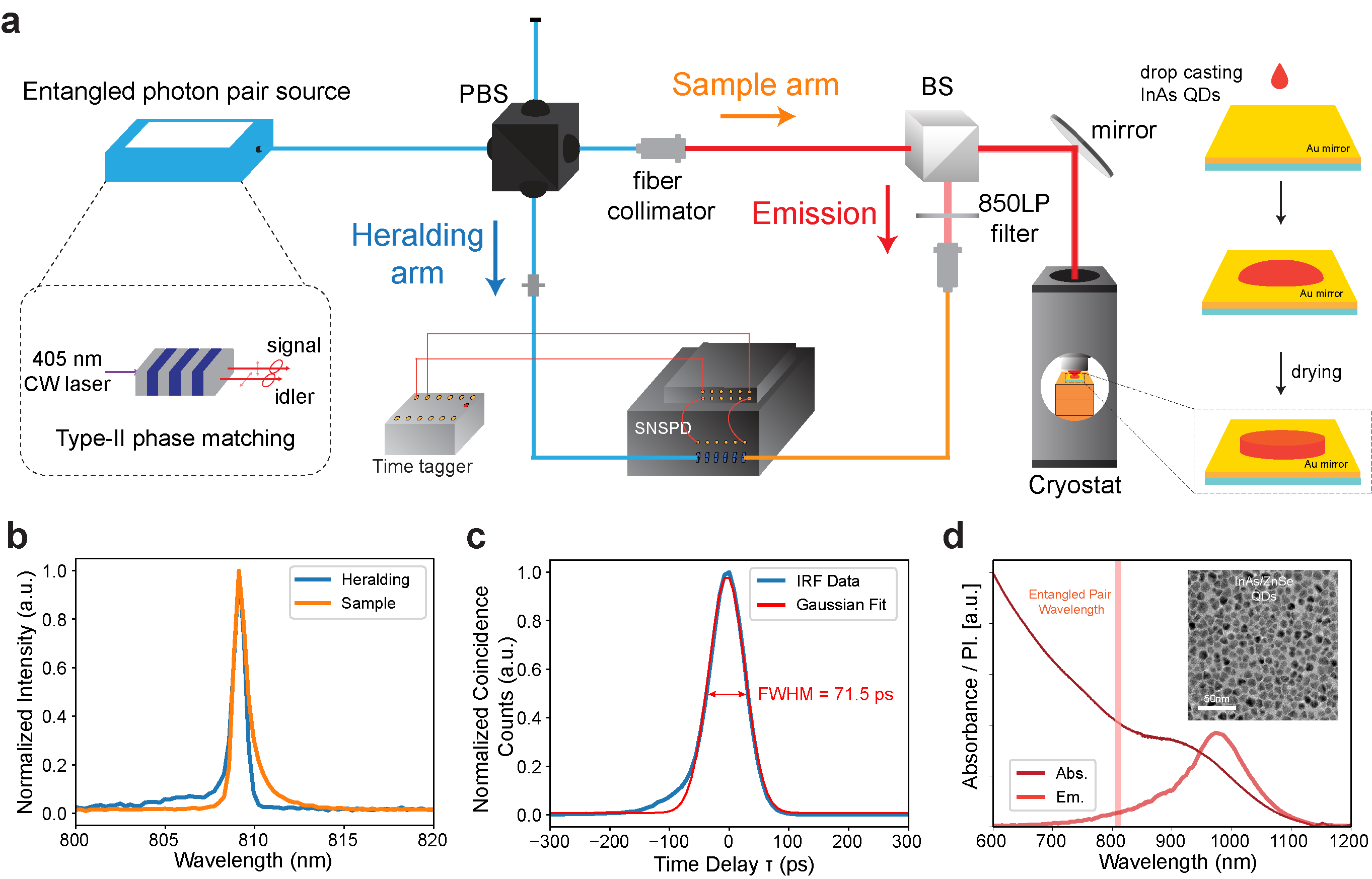}
    \caption{\textbf{Experimental setup, source characteristics, and properties of InAs/ZnSe nanocrystals.} (a) Experimental setup for heralded emission detection and preparation of InAs/ZnSe quantum dot (QD) films. (b) Normalized spectra of the entangled photons from Type-II SPDC used in the heralding and sample arms. (c) The impulse response function (IRF), obtained by replacing the sample with a cover slip coated with a thin gold film and collecting the reflected photons. (d) Ensemble absorption and emission spectra of InAs/ZnSe QDs dispersed in toluene at room temperature. The shaded region marks the twin-pair excitation wavelength in HED. For clarity, the excitation line is drawn with exaggerated width and does not represent the actual narrow bandwidth. Inset: representative transmission electron microscopy (TEM) images of InAs/ZnSe QDs. Scale bar: 30 nm.}
    \label{fig1}
\end{figure}

We developed an improved HED apparatus capable of operating over variable solid-state sample temperatures, as shown in Fig. \ref{fig1}a. A detailed description of the setup components is provided in the Supporting Information. Briefly, entangled photon pairs are generated via SPDC by pumping a periodically poled potassium titanyl phosphate (ppKTP) crystal with a 405 nm CW laser under the type-II phase-matching condition. The generated pairs exhibit a bandwidth of $\sim$2 nm with spectral overlap consistent with degenerate SPDC (Fig. \ref{fig1}b). A Hong–Ou–Mandel (HOM) visibility of 86\% confirms the high indistinguishability of the photon pairs, in agreement with previous reports (Fig. \ref{figSI_HOM_irf}a) \cite{osorio2013purity}. The two photons are separated by a fiber-coupled polarization beam splitter (PBS), yielding a heralding arm and a sample arm. The heralding photons are directly detected by an SNSPD, while the sample-arm photons are sent through a 50:50 beamsplitter into a cryostat with a 0.9 numerical aperature (NA) cryo-objective to excite the sample. The resulting emission is collected by the same objective, filtered from the excitation photons, and coupled via a fiber collimator into a second SNSPD. The use of SNSPDs is a notable feature of our apparatus due to their high photo-detection efficiency in the NIR ($\sim$80\%), minimal dark counts ($<$10 Hz), and low dead time ($\leq$35 ns). Their additionally low timing jitter (31 ps and 50 ps) provides an overall narrow ($\sim$72 ps) instrument response function (IRF) for coincidence counts without sample (Fig. \ref{fig1}c). The IRF width is mainly determined by the convolution of the timing jitter of the SPDC source, SNSPDs, and the single-photon counting module (19 ps).

Colloidal InAs/ZnSe core–shell QDs were selected as the model system for demonstrating HED \cite{li2025shell}. These nanocrystals are promising NIR emitters \cite{jalali2022indium} and have been widely incorporated into photonic devices such as photoconductors \cite{zhao2019general}, phototransistors \cite{kim2023ultrasensitive}, and photodiodes \cite{choi2021ligand, sun2022fast}. A key advantage of colloidal InAs QDs is their compatibility with planar photonic architectures: they can be directly deposited onto a variety of substrates and integrated with on-chip photonic structures \cite{elshaari2020hybrid}, offering a practical route toward scalable quantum photonic technologies relevant to optical communication and quantum information processing.

Our synthesized type-I heterostructured InAs/ZnSe core–shell nanocrystals display high (approximately 59\%) ensemble photoluminescence quantum yield (PLQY) in solution, consistent with previously reported values for this material system \cite{li2025shell}. This PLQY is substantially higher than of organic dyes in the NIR \cite{rurack2011fluorescence}, rendering NIR-QDs a natural choice for HED with high signal strengths. Room temperature QDs ensemble absorption and emission spectra reveal a low prominence excitonic shoulder around 905 nm based on Elliott model fitting (Fig. \ref{figSI_s1_char}g) and a redshifted, broad emission around 984 nm, shown in Fig. \ref{fig1}d. We attribute the absence of narrower features to inhomogeneous broadening due to size- and shape-inhomogeneity, which is common even for the currently best InAs QDs \cite{jalali2022indium,gazis2023colloidal}. The transmission electron microscopy (TEM) image, as shown in the inset of Fig. \ref{fig1}d, confirms the tetrahedral QD shape \cite{zhu2023boosting} and a mean particle size of 8.2 $\pm$ 2.2 nm, with a core size of 4.3 nm and shell thickness of 2.0 nm (Fig. \ref{figSI_s1_char}f). This size is only a fraction of the InAs bulk Bohr radius of around 35 nm \cite{kan2003synthesis}, rendering our QDs well within the strong confinement regime, where size dispersion has a high impact on spectral broadening \cite{jalali2022indium}. 

For HED measurements, we prepared emissive QD thin films by drop-casting onto gold-coated coverslips (Fig.~\ref{fig1}a). Compared with conventional spin-coating, drop-casting produces thicker, higher-optical-density films that result in stronger photoluminescence, thereby improving signal (see Supporting Information).

\begin{figure}
    \centering
    \includegraphics[width=15cm]{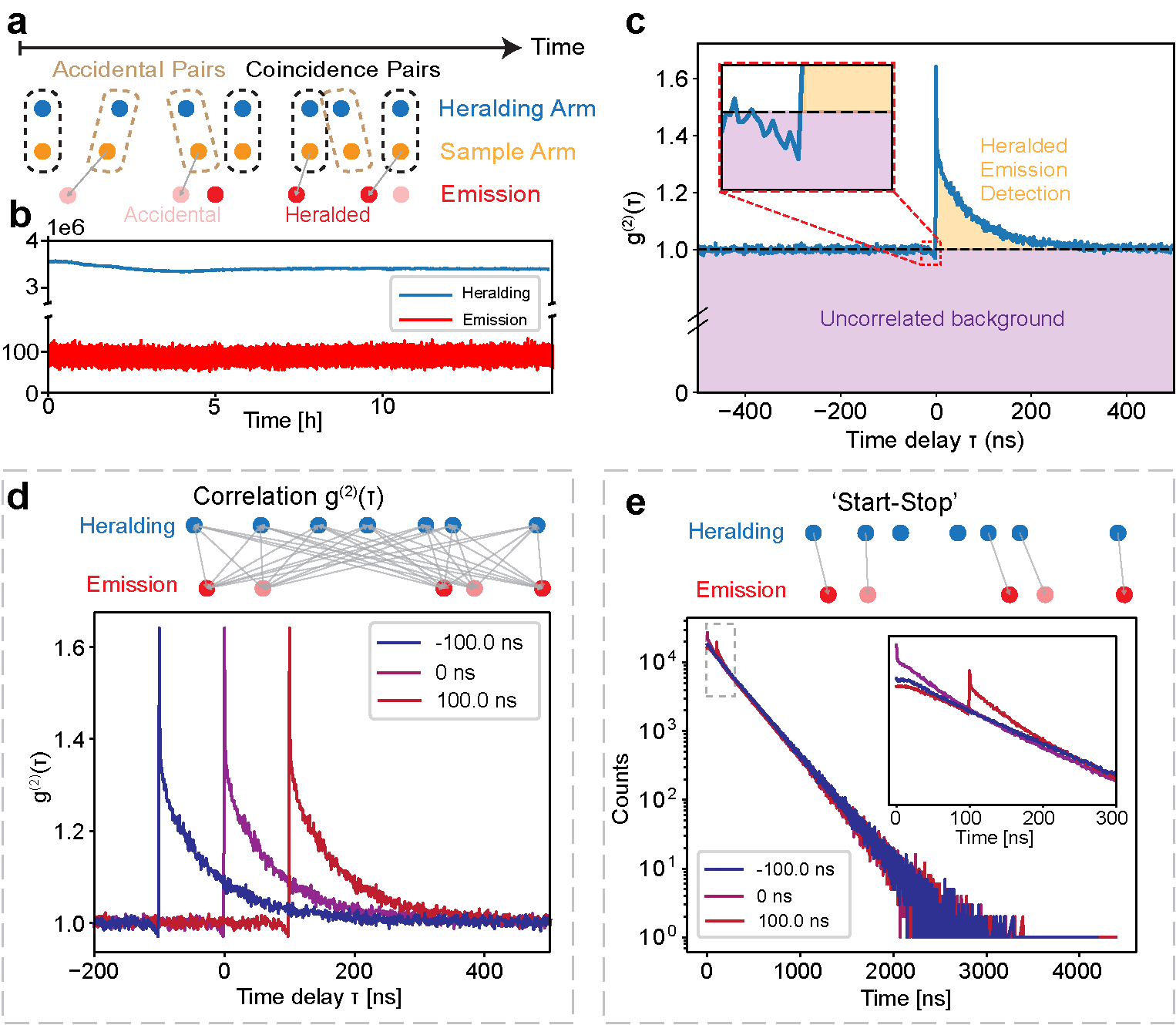}
    \caption{\textbf{Data analyses in heralded emission detection.}  (a) In heralded emission detection (HED), photon pairs are split into a heralding arm (blue) and a sample arm (orange). Pairs can be classified as coincidence pairs, which are time-correlated and produce bunching in the $g^{(2)}(\tau)$, or non-coincidence pairs, which contribute uncorrelated background. The sample emission therefore possesses a similar statistics with emission photon from both twins (red) and accidental pairs (light red). (b) Intensity traces of the heralding arm and the emission signal during a 20h measurement at 4.7 K, demonstrating stable down-conversion over a ten-hour period. (c) Normalized cross-correlation, $g^{(2)}(\tau)$, between the heralding arm and the emission at 4.7 K. The orange-shaded region corresponds to contributions from correlated heralding–emitted photon pairs (entangled two-photon down-conversion), while the purple region represents uncorrelated background with $g^{(2)}(\tau)=1$. (d) Schematic for the $g^{(2)}$ correlation analysis of time-tagged HED data. Delaying one sequence relative to the other only shifts the bunching peak without changing its shape. (e) Schematic of the effect of adopting a 'start–stop' alternative from time-correlated single-photon counting (TCSPC) to HED, introducing distortions.}
    \label{fig2}
\end{figure}

Fig. \ref{fig2} illustrates the concept, stability of emission, and different approaches for analyzing HED data. Three photon “arms” are considered in the measurement: the heralding arm, the sample arm used to excite the sample, and the emission arm, which collects photons emitted by the sample following excitation. SPDC sources are not ideal \cite{ramelow2013highly}; in addition to generating true coincidence photon pairs, non-coincidence pairs are also produced, as shown in Fig. \ref{fig2}a. Consequently, the emitted photons originate from both non-coincidence and coincidence pairs. The former produces accidental heralding-emission coincidences. The sample emission count rates remained stable over the course of the experiment ($\sim$15–18 hours), confirming the sample’s stability under low excitation fluences (see Fig. \ref{fig2}b for 4.7 K). At 4.7 K, the heralding arm consistently recorded an average count rate of $\sim3.2 \times 10^6$ counts per second (cps). Assuming the sample arm exhibits a similar rate, and accounting for losses through optical elements, the number of excitation photons reaching the sample is estimated to be $1.12 \times 10^6$, corresponding to a photon power of $\sim$275 fW. Emitted photons were collected at a rate of $\sim$89 cps (see Supporting Information for details).

We now analyze correlations (see Supporting Information for details) between heralding and emission photons, extending approaches previously applied to molecular dyes at room temperature \cite{harper2023entangled, li2023single, eshun2023fluorescence,gabler2025benchmarking,alvarez2025correlated}. Leveraging the short 72 ps IRF of our HED setup, we resolve the normalized cross-correlation, $g^{(2)}(\tau)$, between emission and heralding photons (Fig. \ref{fig2}c). As expected, a distinct bunching peak $>1$ at $\tau=0$ indicates preservation of the bunched statistics of twin pairs in the heralded emission. The background with $g^{(2)}(\tau)=1$ results from accidental coincidence events. The decay of $g^{(2)}(\tau)>1$ with $\tau$ measures the effective delay between heralded photon absorption and emission, i.e. the spontaneous emission lifetime \cite{harper2023entangled}. We confirm the near-equivalency of HED lifetimes of our QDs and TCSPC, as shown in the Supporting Information. Small deviations are most likely due to differences in the excitation wavelength or small deviations in setup alignments between different excitation sources.
Interestingly, we identify a consistent dip of $g^{(2)}<1$ for $\tau<0$ (Fig. \ref{fig2}c, inset). Such anti-correlations would be consistent with time reversal symmetry breaking, but likely have a more trivial origin.
We assign the origin of this dip to a non-obvious artifact from detector deadtime. Coincidences with $\tau<0$ are from emitted photons leading heralding photons. A leading emitted photon has a larger than random probability of being led by a second, even earlier heralding photon. The detection of this second, earliest, heralding photon turns the heralding arm SNSPD dark for the deadtime ($<$20 ns). The number of following emission/heralding coincidences within the deadtime is thus reduced, producing the dip in $g^{(2)}(\tau<0)$. This artifact may be misinterpreted or may go unnoticed, especially for longer deadtime detectors.
While the above correlation analysis is established, we like to point out that translation of the “start-stop” paradigm used in pulsed-laser TCSPC would require correction in post-processing and is more prone to distortion from pathlength differences. We clarify this in Fig. \ref{fig2}d-e.

The start–stop algorithm selects a photon on one channel as “start” and a photon on the other channel as “stop,” storing their time differences in a histogram. By contrast, the correlation algorithm accumulates all time differences between detection events on the two channels into a histogram, treating every click as both “start” and “stop,” and calculating both positive and negative delays. Using a start–stop scheme in HED can introduce distortions, since the overall number of pairs along the lifetime not only depends on the heralding ratio and entanglement time, but also the waiting time distribution of the accidental pairs. This can produce spurious accidental coincidences that are difficult to separate from the lifetime decay. Timing misalignment between emitted photon and their heralding photon can further deform the decay dynamics (Fig. \ref{fig2}d). The $g^{(2)}$ correlation approach avoids these complications, as it does not impose fixed start/stop assignments on the photons. Any time delay between the heralding and emission arms merely shifts the correlation peak without distorting its decay dynamics (Fig. \ref{fig2}e). Waiting time distributions between accidental counts are naturally factored out in the $g^{(2)}$.

\begin{figure}[ht]
    \centering\includegraphics[width=16cm]{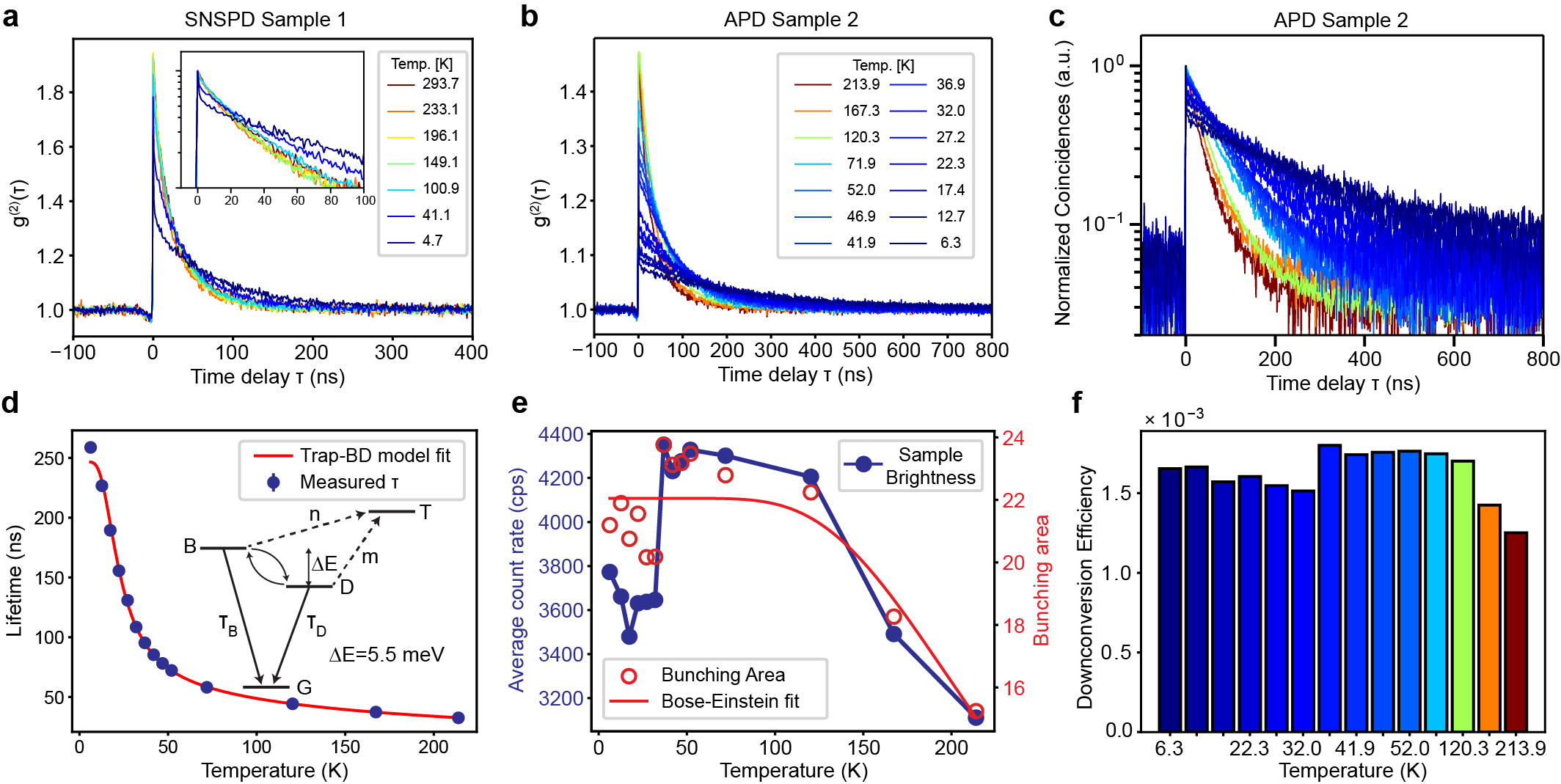}
    \caption{\textbf{Temperature-dependent HED with SNSPD and APD for sample emission detection.} (a) Normalized cross-correlation, $g^{(2)}(\tau)$, at various temperatures, measured by tagging the sample emission with an SNSPD. The bunching peak reflects the time correlations determined by the photoluminescence lifetime of the InAs/ZnSe quantum dots (QDs). The inset shows a zoomed-in view on a logarithmic scale, with $g^{(2)}(\tau)$ normalized so that all peak values are equal, facilitating comparison of the decay dynamics. (b) $g^{(2)}(\tau)$ of a second QD sample measured with an avalanche photodiode (APD). (c) Data from (b) plotted as $G^{(2)}(\tau)$ normalized on a logarithmic scale. (d) Extracted lifetimes as a function of temperature, along with bright–dark model fits. The inset illustrates the model, where G, B, and D denote the ground, bright, and dark state, respectively. (e) Brightness and bunching peak area extracted from $g^{(2)}(\tau)$ data as a function of temperature. The bunching area is fitted with a Bose–Einstein model. (f) External quantum efficiency (EQE) of the twin photon downconversion by the sample at different temperatures, following the trend in brightness.}
    \label{fig4}
\end{figure}

One shortcoming of many SNSPDs is their single-mode fiber access \cite{kim2011efficient}, causing substantial insertion losses. If high count rates rather than minimal timing jitter are desirable in HED, large area APDs are preferred. Fig. \ref{fig4}a–c showcases HED results acquired across different temperatures using both an SNSPD (Fig. \ref{fig4}a) and a large-area APD (Fig. \ref{fig4}b-c). Although the employed APD exhibited high dark counts ($\sim$1848 cps), they contribute only to accidental coincidences and do not affect the shape of the bunching peak (see Supporting Information for details). We consistently observed total sample emission count rates of $\sim$2000 cps when subtracting accidental dark counts. Sample 1, with a quantum dot size of 8.2 nm, and Sample 2, with a size of 6.0 nm, showed similar estimated brightness despite their differing lifetimes. These detected count rates led to an increased external quantum efficiency (EQE) for the film i.e., the ratio of detected emission photons to heralded sample arm photons (Fig. \ref{fig4}f).

To demonstrate the HED analysis under these count rates, we applied it to quantify temperature-dependent lifetimes. The decay dynamics at all temperatures are well described by exponential models containing at least two components, consistent with Sample 1. Specifically, for temperatures $T \leq 52.0$ K, the data were fitted using a biexponential model, where one component captures an initial fast decay and the other represents the slower average lifetime associated with the bright and dark exciton states. At higher temperatures, where the fast component vanishes, a single-exponential fit was sufficient for consistency.

With increasing temperature, the fastest decay component—dominant at low temperatures—gradually diminishes. This behavior is intrinsic to the nanocrystals and not influenced by the gold-coated substrate, as control samples deposited on plain glass slides exhibit the same feature (see Fig.~\ref{figSI_s1_nogold}). The presence of this initial fast decay in all HED traces is consistent with previously reported rapid components in time-resolved photoluminescence measurements of InSb \cite{liu2012colloidal, busatto2020luminescent} and InAs/CdSe \cite{bischof2015origin} nanocrystals. Such fast decays are typically attributed to either nonradiative recombination \cite{liu2012colloidal, busatto2020luminescent} or relaxation from the bright to dark exciton state \cite{bischof2015origin}. Analysis of the temperature dependence of the fast component’s fractional amplitude (see Supporting Information) reveals two activation energies, 2.96~meV and 15.99~meV, which can be ascribed respectively to acoustic-phonon–mediated relaxation or shallow surface/ligand traps, and to deeper trap states or defect-assisted nonradiative decay.

The slower decay component is attributed to thermally activated dark-to-bright exciton population \cite{de2006size, oron2009universal}, reflecting the effective lifetime of excitons in thermal equilibrium. The average lifetime extracted from these fits can be modeled as arising from two thermally coupled excitonic states separated by an energy splitting $\Delta E$, together with an additional trap-assisted decay channel \cite{murphy2016temperature, gaponenko2010temperature}. The bright state has a shorter lifetime $\tau_B$, while the dark state has a longer lifetime $\tau_D$, consistent with the bright–dark exciton framework \cite{murphy2016temperature, gaponenko2010temperature}:

\begin{equation}
\tau_\mathrm{avg}^{-1} = \frac{a \tau_B^{-1} + \tau_D^{-1} + k_0 \left[ a (b-1)^{-n} + (b-1)^{-m} \right]}{1 + a},
\end{equation}

where
\begin{align}
a = e^{-\Delta E / (k_B T)}, \quad b = e^{E_\mathrm{ph} / (k_B T)}.
\end{align}
Here, $k_0$ denotes the trap-assisted decay rate, $E_\mathrm{ph}$ the relevant phonon energy, and $n$ and $m$ the phonon interaction exponents ($m > n$).

Fitting results (Fig.~\ref{fig4}d) yield $\tau_D \approx 246.6$ ns, $\tau_B \approx 21.4$ ns, $\Delta E \approx 5.5$ meV, $E_\mathrm{ph} \approx 23.8$ meV, $1/k_0 \approx 1.03$ ns, $m \approx 1.03$, and $n \approx 0.79$. The extracted $\tau_D$ is shorter than the microsecond-scale dark-state lifetimes typically observed in InAs QDs \cite{oron2009universal}, but comparable to those of InP/ZnSe QDs \cite{brodu2018exciton}. The obtained $\Delta E$ agrees well with reported bright–dark energy splittings of 2–9 meV in other III–V systems \cite{brodu2018exciton, chandrasekaran2023exciton}. The derived phonon energy $E_\mathrm{ph}$ also aligns with optical phonon energies (24–32 meV) of the ZnSe shell \cite{brodu2018exciton}. The small deviations observed at 6.3 K may result from incomplete isolation of the fastest decay component during fitting or the need for a more comprehensive model that explicitly includes fast relaxation dynamics between bright and dark states.

Finally, Fig.~\ref{fig4}e illustrates how both the sample brightness and the total area under the $g^{(2)}(\tau)$ bunching peak (Fig.~\ref{fig4}b) decrease with increasing temperature. The reduction in brightness follows a Bose–Einstein–like dependence on phonon population, with a fitted characteristic energy of $\approx 65$ meV—consistent with phonon-mediated activation of nonradiative decay in III–V QDs \cite{brodu2018exciton, brodu2020exciton}. The reduced bunching peak area can be explained by the increased influence of detector dark counts as QD emission weakens at higher temperature, leading to a lower signal-to-dark-count ratio and diminished bunching contrast. This effect is less pronounced in Sample 1, where SNSPDs with lower dark count rates were used.

\section{Discussion}
Advancing quantum spectroscopy will require both improved apparatus capabilities and broader material platforms. This study contributes by identifying colloidal InAs/ZnSe QDs as bright NIR model systems and by detailing heralding analysis, detector tradeoffs, and unexpected artifact sources to guide future improvements. The main limitation in quantum spectroscopy generally is the low coincidence rate after sample interaction \cite{harper2023entangled, li2023single}. Pinpointing losses to external quantum efficiency across setups \cite{kim2011efficient} and materials is a logical next step. Given the tunability and integrability of QDs as photonic building blocks \cite{ibanez2025prospects}, their targeted optimization could deliver disproportionate gains.

Table \ref{table3} compares key parameters and achieved coincidence rates of all HED studies known to the authors, all but \textit{Eshun et al.} reported low - at most around 120 -coincidence counts per second. One limitation of organic molecules is their generally decreasing luminescent quantum yield (QY) with decreasing emission energy. Few dyes in the near- and shortwave infrared have appreciable QY \cite{rurack2011fluorescence}, naturally limiting the coincidence rate for molecules excitable with established sources typically operating in the NIR. \textit{Eshun et al.} is a notable exception, producing 488–550 nm photons in the sample arm of non-degenerate SPDC pairs, enabling excitation of Rhodamine 6G with an appreciable quantum yield (95\%). While this emission efficiency is certainly contributing to the suggested coincidence rate, we point out that there is currently no agreed-upon best practice for calculating these rates.

Our NIR QDs, when integrated with our detection setup, represent the brightest NIR emitters applied to HED to date. However, maximizing the thin-film PLQY alone is insufficient to achieve the highest HED response, as the detected brightness is also influenced by film thickness, morphology, and sample configuration. Thicker films can lead to reabsorption losses, while inhomogeneous films may cause scattering or trapping of emission, reducing photon out-coupling. In addition, the number of QDs within the focal volume plays a critical role: unlike smaller organic dyes, the larger physical size of QDs lowers the probability of absorbing all incident photons, even though their PLQY is typically higher. Detection using single-mode fiber–coupled SNSPDs introduces further constraints. While these fibers provide spatially localized information from the film, they inherently limit collection efficiency because the QD emission is spatially inhomogeneous across the thin film \cite{li2018effect, kim2011efficient}. Therefore, achieving optimal HED brightness requires simultaneous optimization of absorption and emission into Gaussian-like optical modes, minimizing reabsorption in thicker films, and carefully considering QD density and emission distribution. These considerations are crucial for enhancing photon out-coupling and extending the applicability of HED in quantum photonic systems.

In our other configuration—using an SNSPD in the heralding arm and a large-area NIR APD in the emission arm—we achieved high coincidence count rates of $\sim$230 coincidence counts per second (ccps) at low temperatures. This setup maximizes detection efficiency for both herald and emitted photons while avoiding fiber-coupling insertion and losses from the incoherent sample emission into non-Gaussian modes. The trade-off, however, is the APD’s larger timing jitter, which broadens the instrument response function (Fig. \ref{figSI_HOM_irf}b) and reduces the precision of $g^{(2)}(\tau)$. These results highlight the importance of detector choice in entangled-photon spectroscopy and suggest that advances in quantum light sources and detectors will further benefit studies of quantum light–matter interactions \cite{tsao2025enhancing}.

Overall, this study establishes colloidal QDs as a versatile solid-state platform for exploring entangled-photon light–matter interactions and for developing twin-photon detectors and entangled-photon downconversion devices. Furthermore, it provides a reliable approach to study photon-bunching dynamics modulated by emitter lifetimes. While our current data do not resolve effects of prolonged sample coherences on entangled-photon absorption, future phase-sensitive experiments—such as Franson-type interferometry \cite{fasel2005energy}—could directly probe entanglement preservation and coherence dynamics. The high brightness of our NIR QDs demonstrated here not only facilitates such measurements but also opens avenues for their integration into practical quantum photonic systems, including quantum communication, quantum metrology, and on-chip entangled-photon sources. Taken together, these results highlight the promise of colloidal QDs as both a fundamental platform for studying quantum light–matter interactions and a practical building block for emerging quantum technologies.

\begin{table}[ht!]
\centering
\caption{Summary of HED work to date}
\label{table3}
\resizebox{\textwidth}{!}{%
\begin{tabular}{l p{1.8cm} p{1.8cm} p{1.8cm} p{2cm} p{2cm} p{2cm} p{2cm}}
\toprule
Work & Sample & Lifetime (ns) & Pump & EPP (ccps) & Excitation (cps) & Emission (cps) & Coincidences (ccps) \\
\midrule
Harper et al. \cite{harper2023entangled} & ICG  &  0.62 & CW & -- & $1\times10^{5}$ & 860 & 6.5 \\
Li et al. \cite{li2023single}            & LH2  & 1.20 & Pulsed & $2.7\times10^{5}$ & $9.13\times10^{5}$ & -- & 121 \\
Eshun et al. \cite{eshun2023fluorescence}* & R6G  & 3.99 & CW & $1.6\times10^{6}$ & -- & -- & 23000 \\
Gäbler et al. \cite{gabler2025benchmarking} & IR-140 & 0.93 & CW & -- & -- & -- & -- \\
Alvarez-Mendoza et al. \cite{alvarez2025correlated} & LH2 & 1.14 & CW & $2\times10^{5}$ & -- & -- & -- \\
This work** & InAs/ZnSe QDs & 40–250 & CW & $1.6\times10^{5}$ & $1.1\times10^{6}$ & 20–4200 & 4–270 \\
\bottomrule
\end{tabular}%
}
\parbox{\textwidth}{\scriptsize
Note: EPP refers to the coincidence counts per second (ccps) for entangled photon pairs. Coincidences represent counts detected between the heralding and emission arms. For comparison with the literature, we report the maximum recorded values of these counts.\\
*The definitions of EPP and coincidences in some previous works may differ.\\
**Approximations or ranges are provided for each parameter to reflect variations under different experimental conditions.
}
\end{table}

\section{Conclusions}
We have demonstrated heralded emission detection (HED) under entangled-photon excitation in InAs/ZnSe quantum dot films at cryogenic temperatures, establishing a solid-state platform for this spectroscopy. By benchmarking against TCSPC, we show that HED reliably recovers excitonic lifetimes and bright–dark state dynamics, validating its extension beyond molecular systems. 
Our results highlight both opportunities and limits. QDs provide brightness, tunability, and device compatibility that make them promising for quantum-light spectroscopy. While this work produced higher coincidence rates than all but one previous work \cite{eshun2023fluorescence}, and the highest in the NIR spectral range, the observed coincidence rates are still modest. The demonstrated integration of QDs with optical cavities and waveguides presents a new control knob in quantum light spectroscopy, positioning QDs as a versatile model system.

\section{Methods}
\subsection{Synthesis of InAs/ZnSe QDs}
InAs/ZnSe QDs were synthesized following previously reported methods \cite{asor2023zn,li2025shell,tamang2016tuning}. Both InAs nanoclusters and InAs QDs were prepared using the same precursor.
\subsubsection{Materials} 
Indium(III) acetate (\ce{In(Ac)3}, 99.99\% trace metals basis), oleic acid (OA, 90\%), 1-octadecene (90\%, ODE), tris(tri-methylsilyl) arsine (99\%, \ce{(TMS)3As}), zinc stearate (\ce{Zn(St)2}, 10-12\% Zn basis), squalane (90\%), selenium (powder, 100 mesh, 99.99\% trace metals basis), toluene (anhydrous, 99.8\%), ethanol (anhydrous, 99.8\%)  were purchased from Sigma-Aldrich. All precursors and solvents were used without further purification. All solvents were stored inside an \ce{N2}-filled glovebox. 

\subsubsection{Synthesis of InAs Clusters} 
To prepare the InAs clusters solution,  12 mmol \ce{In(Ac)3} and 36 mmol OA were mixed with 60 mL of ODE. The mixture was degassed at 120\textdegree{}C under vacuum for 90 min, then cooled down to room temperature under an Argon atmosphere. Separately, 1.6 g \ce{(TMSi)3As} was mixed with 12 mL of dry ODE in the glovebox. The As solution was injected at room temperature into the indium-oleate solution with constant stirring. The reaction mixture was then heated slowly to 80\textdegree{}C until the solution turned dark red, indicating the formation of InAs nanoclusters.  The resulting InAs nanoclusters were cooled down to room temperature and stored under an inert atmosphere in the glovebox. 

\subsubsection{Synthesis of InAs QDs}
For the synthesis of InAs QDs, 2 mmol of \ce{In(Ac)3} and 6 mmol of OA were mixed together with 5 mL of ODE and left under vacuum at 120\textdegree{}C for 2 h. The atmosphere was then switched to argon, and the temperature was raised to 300\textdegree{}C. Meanwhile, 0.24 g \ce{(TMSi)3As} was mixed with 2 mL of dry ODE in the glovebox. The As solution was swiftly injected into the hot indium-oleate solution at 300\textdegree{}C under constant stirring to initiate nucleation. The temperature was then lowered to 285\textdegree{}C and held constant for 20 min. During this period, aliquots were taken, and their absorption spectrum was measured. Once the first excitonic absorption peak of the seeds reached approximately 750 nm, the InAs nanoclusters solution was added continuously at a constant rate of 2 \text{mL}·\text{h}$^{-1}$. During this reaction aliquots were taken at constant intervals to evaluate the growth and size focusing of the nanocrystals by absorption measurements. After the reaction, the mixture was cooled to room temperature and transferred into the glovebox. The resulting QDs were purified through repeated precipitation-centrifugation-redispersion cycles using dry toluene and ethanol as solvent and antisolvent, respectively. The purified InAs QDs were finally dispersed in toluene and stored in the glovebox for further use.

\subsubsection{Synthesis of ZnSe shell} 
To synthesize InAs/ZnSe core/shell QDs, 0.2 mmol of \ce{Zn(St)2}, 3 ml squalane and the purified InAs core QDs in toluene were loaded in a 50 ml three-neck flask and degassed at 120\textdegree{}C under vacuum for 1 h. After purging with argon, the reaction mixture was heated up to 300\textdegree{}C. The Se suspension (Se-SUS, 0.2 mL, 0.5 M) was added into the solution and held at that temperature for 30 min. The temperature increased to 340 °C under argon. Maintain at this temperature for 30 minutes. Then, 0.4 mL of 0.4 M \ce{Zn(St)2} suspension and 0.5 mL of 0.2 M Se-SUS solution were separately injected dropwise in the flask every 30 min. Aliquots were taken to monitor the shell growth progress. The obtained InAs/ZnSe QDs were purified with toluene and ethanol and dispersed in toluene. All the washing procedures were performed in the glovebox.

\subsection{InAs/ZnSe QDs characterization}
The measurements of UV-vis absorption were performed on a Jasco V-570 UV-vis-NIR spectrophotometer. Photoluminescence spectra and photoluminescence quantum yield was measured by a Hamamatsu Absolute PL quantum yield spectrometer C11347-11 and Edinburgh FLS920 fluorescence spectrometer. Transmission electron microscopy (TEM) images of the InAs/ZnSe QDs were acquired using a Tecnai 12 TEM. Scanning electron microscopy (SEM) images of the QD films were acquired using a Zeiss XB 550 FIB-SEM at 0.7 kV with a working distance of 4 mm.

\section*{Acknowledgments}
We thank Tim Rambo and Ryan Wilson for valuable discussions regarding instrumentation. This work was supported by the College of Chemistry at the University of California, Berkeley. Work at the Molecular Foundry was supported by the Office of Science, Office of Basic Energy Sciences, of the U.S. Department of Energy under Contract No. DE-AC02-05CH11231. Partial support was also provided by the Israel Science Foundation through the MAPATS program (U.B., Grant No.~2655/23).


\section*{Supplemental document}
The data supporting the findings of this study are available within the article and its Supplementary Information.

\section*{References}



\printbibliography[heading=none]
\end{refsection}

\clearpage
\newpage

\begin{refsection}
\section*{Supplementary materials}

\renewcommand{\thesection}{S\arabic{section}}
\renewcommand{\thesubsection}{\thesection.\arabic{subsection}}
\renewcommand{\thesubsubsection}{\thesubsection.\arabic{subsubsection}}
\renewcommand{\theequation}{S\arabic{equation}}
\renewcommand{\thetable}{S\arabic{table}}
\renewcommand{\thefigure}{S\arabic{figure}}

\setcounter{section}{0} 
\setcounter{subsection}{0}  
\setcounter{equation}{0}
\setcounter{table}{0}
\setcounter{figure}{0}

\section{The detailed configuration of the setup}  
\subsection{Heraled Emission Detection (HED)}
\begin{figure}[h]
    \centering
    \includegraphics[width=15cm]{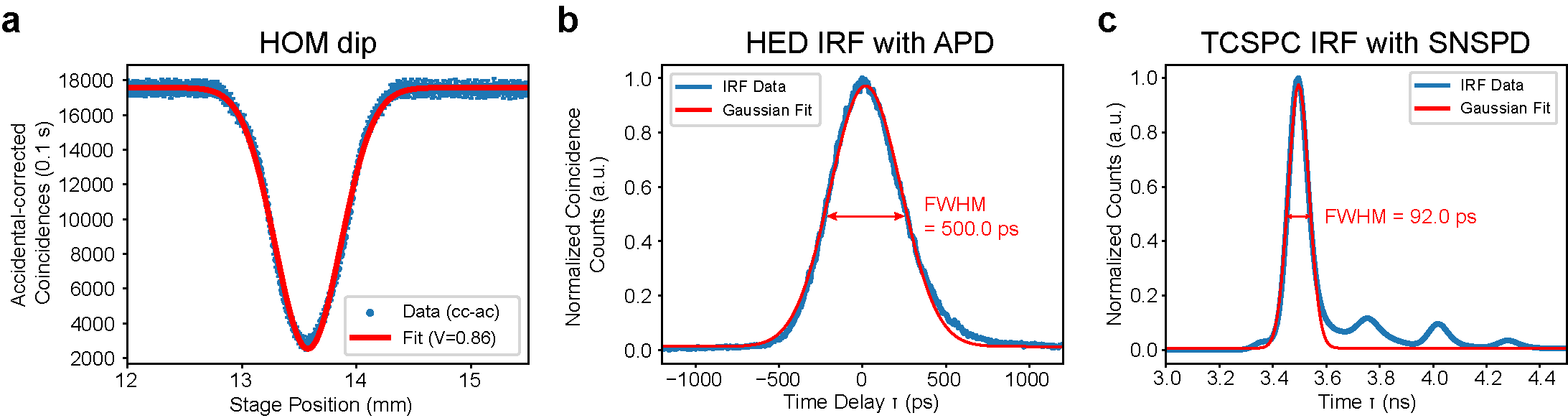}
    \caption{\textbf{Characterization of the setups.} 
    (a) Hong–Ou–Mandel (HOM) dip of the entangled photon pairs, demonstrating their indistinguishability, with a visibility of approximately 86\%. 
    (b) Impulse response function (IRF) of the HED system with an APD detector in the emission arm.
    (c) IRF of the TCSPC system with an SNSPD detector in the emission arm.}
    \label{figSI_HOM_irf}
\end{figure}

The entangled photon pair source (customized QES 2.4, Qubitekk) consists of a 405 nm continuous-wave (CW) laser that pumps a periodically poled potassium titanyl phosphate (ppKTP) crystal housed in a temperature-controlled oven, allowing for tuning of the photon wavelength and degree of degeneracy. The indistinguishability of photon pairs has been shown in Fig. \ref{figSI_HOM_irf}a. Entangled photon pairs are generated via spontaneous parametric down-conversion (SPDC), after which the residual pump light is filtered out. The entangled photons are then coupled into an optical fiber. A polarization-maintaining, fiber-coupled polarizing beam splitter (PBS) with FC/PC connectors (Qubitekk) is connected to the source output to separate the signal and idler photons. One output of the PBS is routed via an FC/PC sleeve (ADAFC2, Thorlabs) to a single-mode fiber patch cable (780HP, Diamond), which is terminated with an E2000 connector and connected to a superconducting nanowire single-photon detector (SNSPD, Quantum Opus), referred to as the heralding arm.

The other output of the PBS is coupled into a protected-silver-coated reflective fiber collimator (RC02FC-P01, Thorlabs), producing a collimated beam with a $1/e^2$ diameter of approximately 1.8 mm. The beam is launched into free space, passes through an 810 nm bandpass filter (FBH810-10, Thorlabs) and protected-silver mirrors (PF10-03-P01, Thorlabs), and is elevated by a periscope assembly (RS99, Thorlabs) equipped with the same mirror type. It then passes through a 50:50 beam splitter (BS014, Thorlabs) and is directed onto an x–y galvanometric mirror system (GVS012, Thorlabs) controlled by a data acquisition (DAQ) platform (T7, LabJack). The reflected beam is subsequently relayed into a high-numerical-aperture (NA) objective (EC Epiplan-Neofluar 100×/0.9 BD DIC M27, Zeiss) housed in a closed-cycle cryostat (Montana Instruments) using a pair of relay lenses with focal lengths of 150 mm (AC254-150-AB-ML, Thorlabs) and 200 mm (AC254-200-AB-ML, Thorlabs). This configuration, together with the beam divergence, expands the spot to fill the objective’s back aperture, which has a conjugate plane size of approximately 3 mm.

Photoluminescence emission from the sample is collected by the same objective and travels back through the same relay lens pair and galvo system. The emission is separated from the excitation path by the same 50:50 BS and filtered by an 850 nm long-pass filter (FELH0850, Thorlabs) to remove residual excitation photons. The filtered emission is then collected by another fiber coupler (PAF2P-A10B FiberPort, Thorlabs) and routed via single-mode fiber patch (1060HI, Diamond) to a second SNSPD, referred to as the emission arm.

Different fibers and detectors are used for the heralding and emission arms to match their respective wavelength ranges. The Heralding arm uses 780HP fiber, which operates effectively in the 780–970 nm range, and is connected to an SNSPD with a timing jitter of approximately 31 ps, less than 1 dark count per second, and a dead time of 20 ns. The Sample arm uses HI1060 fiber, paired with an SNSPD optimized for the emission wavelength of InAs/ZnSe nanocrystals. This detector has a timing jitter of approximately 50 ps, fewer than 10 dark counts per second, and a dead time of 18 ns.

Detection events from both SNSPDs are time-stamped with 19 ps (8 ps in RMS) resolution using a single-photon counting module (Time Tagger Ultra, Performance Edition, Swabian Instruments). For second-order correlation measurements, the system accumulates the time differences between photon detection events on the two channels into a histogram. Each detection event is considered as both a “start” and a “stop,” allowing the calculation of both positive and negative time delays. A time bin width of 1 ns is chosen to balance the trade-off between signal-to-noise ratio (SNR) and timing resolution.

For the APD-based version of the HED, the short-pass dichroic mirror (FF825-SDI01, Semrock) replaces the PBS output path, and the signal is detected using an APD (id120, IDQ). The impulse response function (IRF) for this setup is shown in Fig. \ref{figSI_HOM_irf}b.

It should be noted that the timing jitters here are specified as full width at half maximum (FWHM).

\begin{table}[h]
    \label{detectors}
    \centering
    \caption{Summary of detectors used in this work.}
    \begin{tabular}{lcccc}
    \hline
    Detector & Deadtime (ns) & dark count (cps) & Timing jitter (ps) & detection efficiency\\
    \hline
    SNSPD (heralding) & $<20$@780 nm & $<10$ & 31 & 83\%@780nm \\
    SNSPD (emission) & $\leq35$@1315nm & $<10$ & 50 & 83\%@1064 nm \\
    APD (emission) & 400 & $\sim$1848 & $<$400 & 10\%@1000 nm\\
    \hline
    \end{tabular}
\end{table}

\subsection{Time-Correlated Single Photon Counting (TCSPC)}
The setup for TCSPC measurements closely follows that used for HED, with the primary difference being the excitation source: a 470 nm pulsed laser (LDH-P-C-470, PicoQuant). The laser repetition rate is reduced to 2.5 MHz using an external controller (PDL 820, PicoQuant) to meet the desired excitation conditions. Along the excitation path, a 500 nm short-pass filter (FESH0500, Thorlabs) is combined with a set of neutral density filters (NE20B, NE20A, and NE10A-A; Thorlabs), reducing the excitation fluence to 6.37 nJ/cm², which cause the exciton number per pulse $\langle N\rangle \ll 1$ (See the following section to discuss the value). All other optical components remain unchanged from the HED setup.

For TCSPC measurements, a start-stop mode is used on the same Time Tagger in place of a second-order correlation algorithm. In this mode, detection events on two channels are recorded as start and stop signals, and the time differences between them are compiled into a histogram. For consistency with HED data, the histogram bin width is set to 1 ns.

\section{QDs and Their Film Characterizations}
\subsection{InAs/ZnSe QDs in solution characterizations}
We investigated two types of InAs/ZnSe quantum dot (QD) samples for HED, employing SNSPDs and APDs on the emission side. The ZnSe shell was introduced to enhance the fluorescence efficiency and environmental stability of the QDs.

\begin{figure}[H]
    \centering
    \includegraphics[width=13cm]{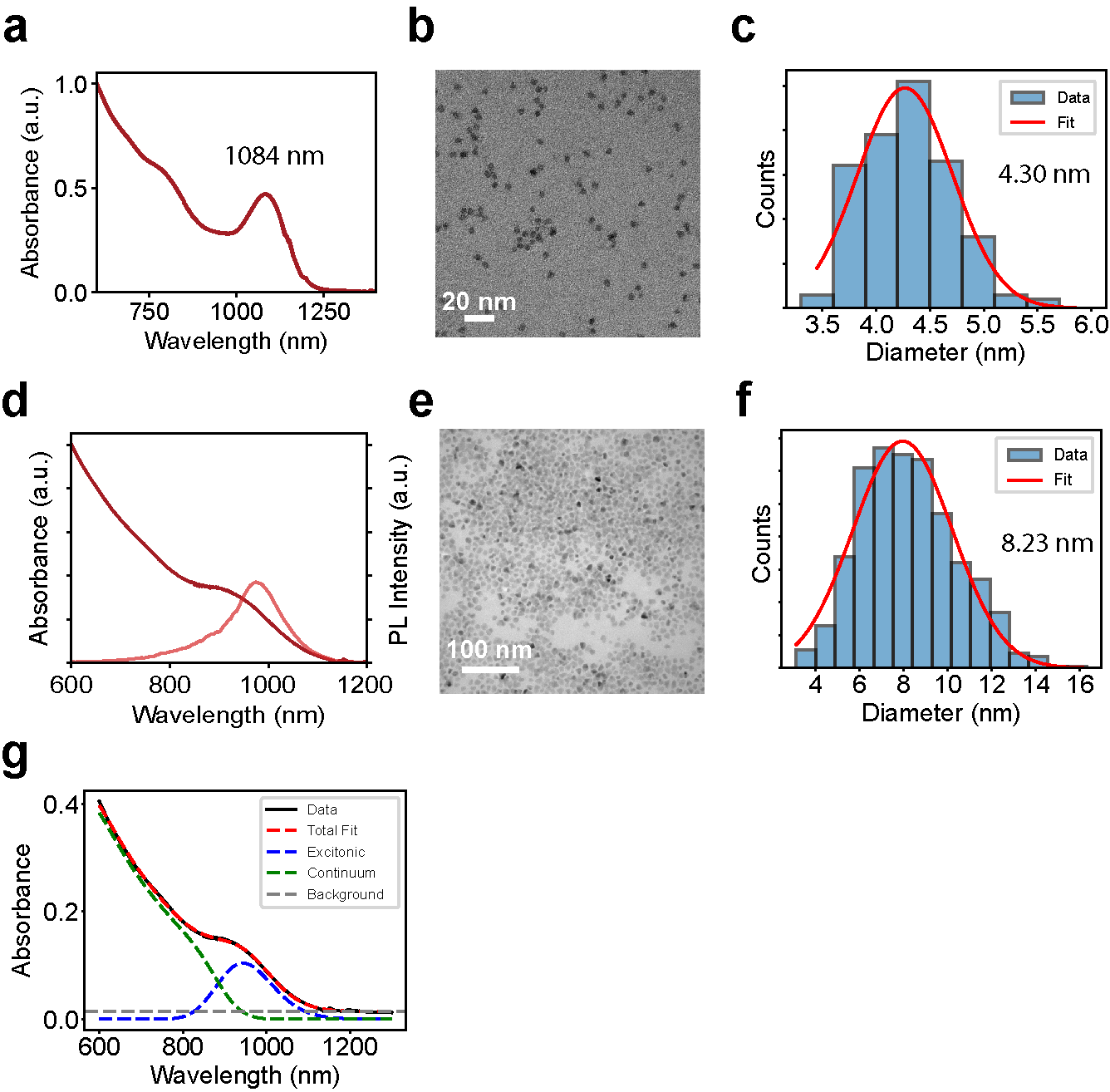}
    \caption{\textbf{Characterization of Sample 1 in solution.} 
    (a) Absorption spectrum of the InAs cores. 
    (b) TEM image of InAs cores. 
    (c) Size distribution of InAs cores extracted from TEM images. 
    (d) Absorption and photoluminescence spectra of InAs/ZnSe QDs. 
    (e) TEM image of InAs/ZnSe QDs. 
    (f) Size distribution of InAs/ZnSe QDs extracted from TEM images. 
    (g) Elliott formula fitting of the absorption spectrum shown in (d).}
    \label{figSI_s1_char}
\end{figure}

For the sample used with the SNSPD (Sample 1), the InAs cores exhibit a distinct first excitonic absorption peak at 1084 nm, corresponding to an average core diameter of approximately 4.30 nm, as determined by transmission electron microscopy (TEM) (Fig. \ref{figSI_s1_char}a-c). Subsequent growth of ZnSe epilayers on the InAs cores results in a type-I InAs/ZnSe core/shell heterostructure. TEM analysis reveals that the average diameter of the InAs/ZnSe QDs is 8.23 nm (Fig. \ref{figSI_s1_char}e-f), indicating a ZnSe shell thickness of about 1.97 nm. Considering that one monolayer of cubic ZnSe corresponds to 0.328 nm, the shell thickness is equivalent to approximately 5.99 monolayers. The corresponding absorption and photoluminescence (PL) spectra are shown in Fig. \ref{figSI_s1_char}d. The PL peak is centered at 984 nm (1.26 eV). To more precisely determine the first excitonic absorption peak, the absorption spectrum of the InAs/ZnSe QDs was analyzed using the Elliott model (Fig. \ref{figSI_s1_char}g), yielding a first excitonic peak at 951.20 nm (1.30 eV). The resulting Stokes shift is approximately 43 meV.

\begin{figure}[H]
    \centering\includegraphics[width=13cm]{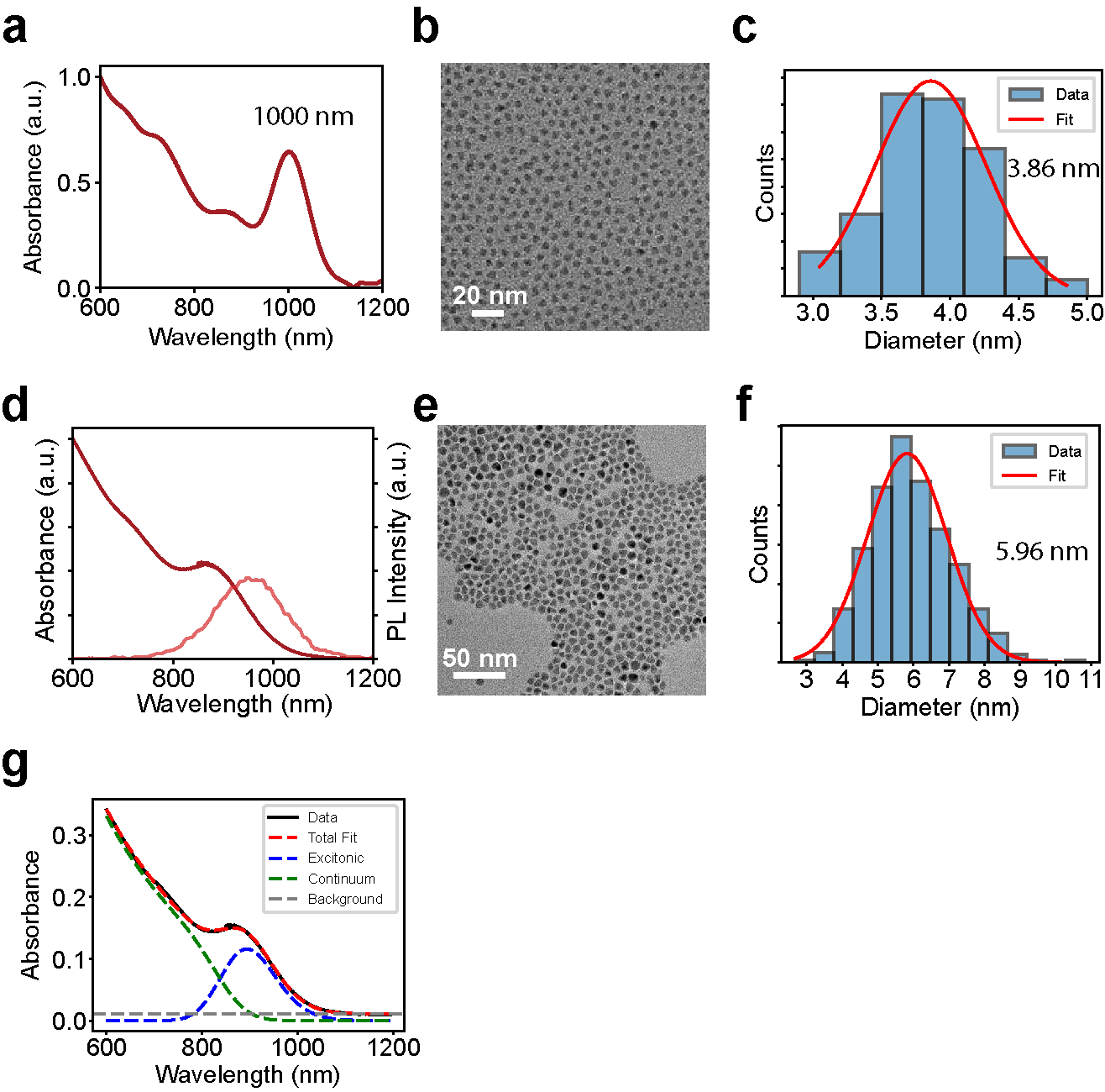}
    \caption{\textbf{Characterization of Sample 2 in solution.} 
    (a) Absorption spectrum of the InAs cores. 
    (b) TEM image of InAs cores. 
    (c) Size distribution of InAs cores extracted from TEM images. 
    (d) Absorption and photoluminescence spectra of InAs/ZnSe QDs. 
    (e) TEM image of InAs/ZnSe QDs. 
    (f) Size distribution of InAs/ZnSe QDs extracted from TEM images. 
    (g) Elliott formula fitting of the absorption spectrum shown in (d).}
    \label{figSI_s2_char}
\end{figure}

For the sample used with the APD (Sample 2), the InAs cores exhibit a first excitonic absorption peak at 1000 nm, also featuring a pronounced sharp transition (Fig. \ref{figSI_s2_char}a). TEM analysis shows an average core diameter of 3.86 nm, as shown in Fig. \ref{figSI_s2_char}b-c. For the corresponding InAs/ZnSe QDs, the average particle diameter of 5.96 nm indicates a ZnSe shell thickness of approximately 3.20 monolayers (Fig. \ref{figSI_s2_char}e). The corresponding absorption and PL spectra are shown in Fig. \ref{figSI_s2_char}d. The PL peak is centered at 957 nm (1.30 eV). Analysis of the absorption spectrum using the Elliott model (Fig. \ref{figSI_s2_char}g) yields a first excitonic peak at 899.98 nm (1.38 eV), giving a Stokes shift of approximately 82 meV.

A summary of the ensemble optical and structural properties of the InAs/ZnSe QDs in solution for this work is provided in Table \ref{InAs_char}.

\begin{table}[h]
    \centering
    \begin{threeparttable}
    \caption{Summary of InAs/ZnSe QDs in this work.}
    \label{InAs_char}
    \begin{tabular}{lcc}
    \hline
    Property & Sample 1 & Sample 2 \\
    \hline
    InAs core diameter (nm) & $4.30 \pm 0.42$ & $3.86 \pm 0.41$ \\
    InAs core absorption peak & 1084 & 1000 \\
    InAs/ZnSe QDs diameter (nm) & $8.23 \pm 2.18$ &  $5.96 \pm 1.13$ \\
    ZnSe shell thickness (nm) & 1.97 &  1.05 \\
    the number of ZnSe monolayers & 5.99 & 3.20 \\
    InAs/ZnSe QDs first excitonic absorption peak (nm/eV)& 951.20/1.30& 899.98/1.38\\
    InAs/ZnSe QDs emission peak (nm/eV) & 984/1.26& 957/1.30\\
    Stokes shift (meV) &43&82\\
    PLQY& 59\% & 71\%\\
    \hline
    \end{tabular}
    \begin{tablenotes}[flushleft]
        \footnotesize
        \item Note: Uncertainties represent one standard deviation. 
    \end{tablenotes}
    \end{threeparttable}
\end{table}

\subsection{InAs/ZnSe QDs film fabrication and characterizations}
Spin-coated quantum dot (QD) films typically form thin layers ($\sim$30–100 nm) \cite{song2018energy, kirmani2020optimizing}, which contain only a limited number of QDs within the optical focal volume (Rayleigh length × focus spot size), thereby reducing photon absorption and leading to weak heralded-emission detection (HED) signals, even for materials with high photoluminescence quantum yield (PLQY). One common strategy to increase film thickness is to incorporate polymers such as polystyrene (PS) or poly(methyl methacrylate) (PMMA), but achieving thicknesses comparable to the Rayleigh length requires high polymer concentrations. This not only reduces the number of QDs in the focal volume and the overall emission intensity, but can also introduce quenching effects. Layer-by-layer (LbL) deposition, in which QDs are sequentially coated and ligand-exchanged, can produce multilayered films \cite{chernomordik2017quantum}, yet often at the expense of reduced PLQY and additional complexity \cite{kirmani2020optimizing}.

In this work, we first evaluated the effect of polymer incorporation on film PLQY using InAs/ZnSe QDs ($\sim$40\% PLQY in toluene). Films were prepared at a fixed QD concentration with varying polymer content and identical drop-cast volumes. Films without polymer retained PLQY comparable to the solution, whereas 10~wt\% polymer reduced PLQY to $\sim$15\%, consistent with the expected decrease in QD density within the focal volume. Based on these results, polymer matrices were excluded from the optimized film formulation.

\begin{figure}[H]
    \centering
    \includegraphics[width=15cm]{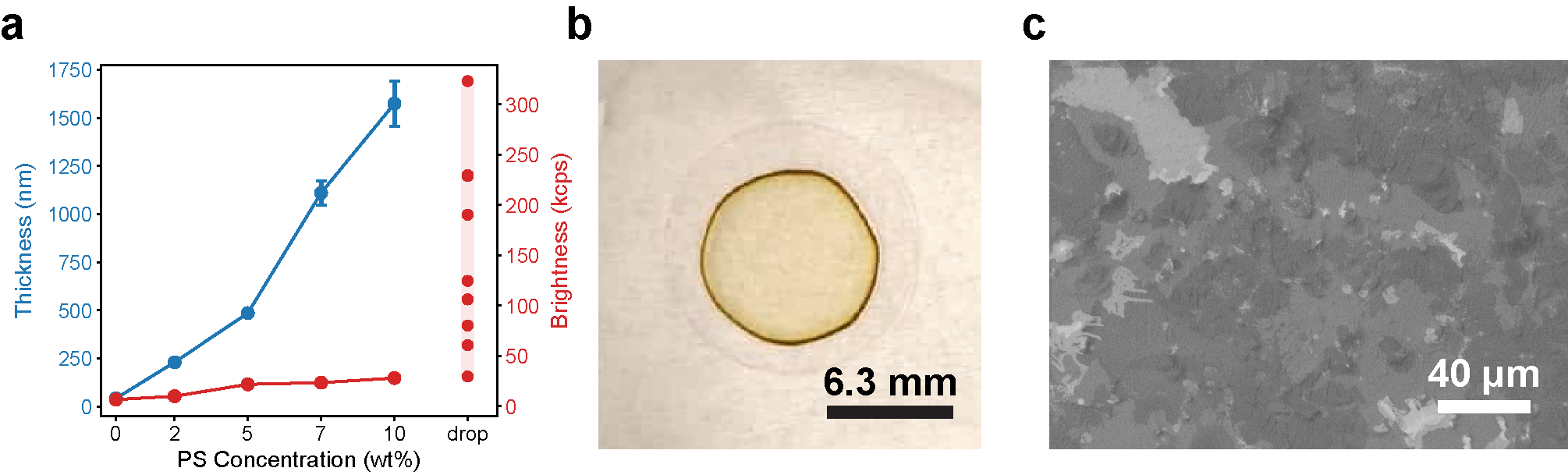}
    \caption{\textbf{Characterization of InAs/ZnSe QD films.} 
    (a) Film thickness and brightness as a function of PS concentration. The point labeled ``drop'' corresponds to the drop-cast film. Due to film inhomogeneity, the eight measurement points show significant variation.
    (b) Photograph of the drop-cast film on a plain glass substrate (without gold mirror) for improved contrast. 
    (c) Top-view SEM image of the InAs QD film. 
    }
    \label{figSI_InAs_film}
\end{figure}

Next, we explored methods to increase film thickness. Spin-coating, even with polymer additives, produced films thinner and dimmer than drop-cast films (Fig.~\ref{figSI_InAs_film}a). LbL deposition also decreased PLQY after four cycles ($\sim$15\%), likely due to nonradiative losses or quenching. By contrast, simple drop-casting of pure QD solutions produced thicker films with PLQY values comparable to the solution phase (Fig.~\ref{figSI_InAs_film}b). While drop-cast films exhibit slight local variations in morphology and coverage (Fig.~\ref{figSI_InAs_film}c), the bright regions provide sufficient signal for HED measurements, making this approach optimal for our study.

\section{Supplementary Analysis of Exciton Population and Elliott Modeling}
\subsection{Estimation of average excitons per pulse}

The average number of excitons generated per pulse in the InAs/ZnSe quantum dots was estimated from the absorption cross section of the InAs core and the excitation fluence. The quantum dots have a core diameter of 4.30~nm and an overall diameter of 8.23~nm due to the ZnSe shell. The absorption cross section at the excitation wavelength of 470~nm was approximated as $\sigma = 6.16 \times 10^{-15}~\mathrm{cm^2}$ based on the core size \cite{yu2005absorption}. The ZnSe shell induces a blue shift of the first excitonic absorption from 1084~nm to 905~nm, slightly reducing the absorption cross section relative to the bare core. Excitation was provided by a pulsed laser with a fluence of 6.37~nJ/cm$^2$ ($1.51 \times 10^{10}$ photons/cm$^2$). Using
\[
\langle N \rangle = \sigma \cdot F,
\]
where $\sigma$ is the absorption cross section and $F$ is the photon flux per pulse, the maximum estimated average number of excitons per pulse is approximately $9.30 \times 10^{-5}$, that is, $\langle N \rangle  \ll 1$. This low excitation regime ensures predominantly single-exciton conditions, which is crucial for reliable photon-correlation measurements and minimizes multi-exciton generation.

\subsection{Absorption spectrum modeling using the Elliott formula}

The absorption spectrum near the band edge was modeled using an extended Elliott formula that accounts for both discrete excitonic transitions and Coulomb-enhanced continuum absorption \cite{sestu2015absorption,yang2016large}. The total absorption coefficient is expressed as

\begin{equation}
	\nonumber
	\begin{aligned}
A(E)&\propto \mu_{cv}^2\sqrt{E_B}\left(\sum_n A_{nx}+A_c\right)\\
&=A_0\left[\sum_{n}\frac{2E_B}{n^3} g\left(E-(E_g-E_B/n^2),\sigma_{ex}\right)+\int_{E_g}^\infty \frac{1+\alpha(E'-E_g)}{1-e^{-2\pi\sqrt{\frac{E_B}{E'-E_g}}}}g(E-E',\sigma_c)dE'\right]+C,
\end{aligned}
\end{equation}
where $A_0$ is a proportionality constant related to the transition dipole moment $\mu_{cv}$; $E_B$ is the exciton binding energy; $E_g$ is the bandgap; and $C$ is a constant offset accounting for the baseline. The term $\alpha(E-E_g)$ accounts for the nonparabolic dispersion of the continuum states, with $\alpha$ being a fitting constant. Both the excitonic and continuum contributions are convoluted with a phenomenological broadening function $g(E,\sigma)$—in this case, Gaussians with standard deviation of $\sigma_{ex}$ and $\sigma_{c}$, respectively—to account for the inhomogeneous broadening.

In the fitting procedure, the parameters $A_0$, $E_g$, $E_B$, $\sigma_{ex}$,  $\alpha$, $\sigma_c$, and $C$ were optimized using a nonlinear least-squares routine. The position of the first excitonic absorption peak, corresponding to the $n=1$ bound state, was then obtained from the fitted parameters as

\begin{equation}
E_1 = E_g - E_B,
\qquad
\lambda_1 = \frac{h c}{E_1}.
\label{eq:first_exciton}
\end{equation}

\section{HED Results and Analysis}
\subsection{Coincidence rates}
To obtain the true coincidence counts from the measured data, it is necessary to subtract the accidental coincidences, which originate from uncorrelated photon detection events occurring within the same coincidence window. The accidental coincidence rate can be estimated as \cite{pearson2010hands, bjurlin2025versatile}
\begin{equation}
R_{\mathrm{acc}} = R_H R_E \tau_c,
\label{eq:accrate}
\end{equation}
where $R_H$ and $R_E$ are the photon detection count rates in the heralding and emission channels, respectively, and $\tau_c$ is the coincidence window, defined as the temporal interval within which detection events are considered coincident. Subtracting accidental counts based on $R_\mathrm{acc}$ from the raw coincidence histogram yields the accidental-corrected coincidence counts, which reflect the true photon-pair correlations between the heralding and emission photons.

\begin{figure}[H]
\centering
\includegraphics[width=15cm]{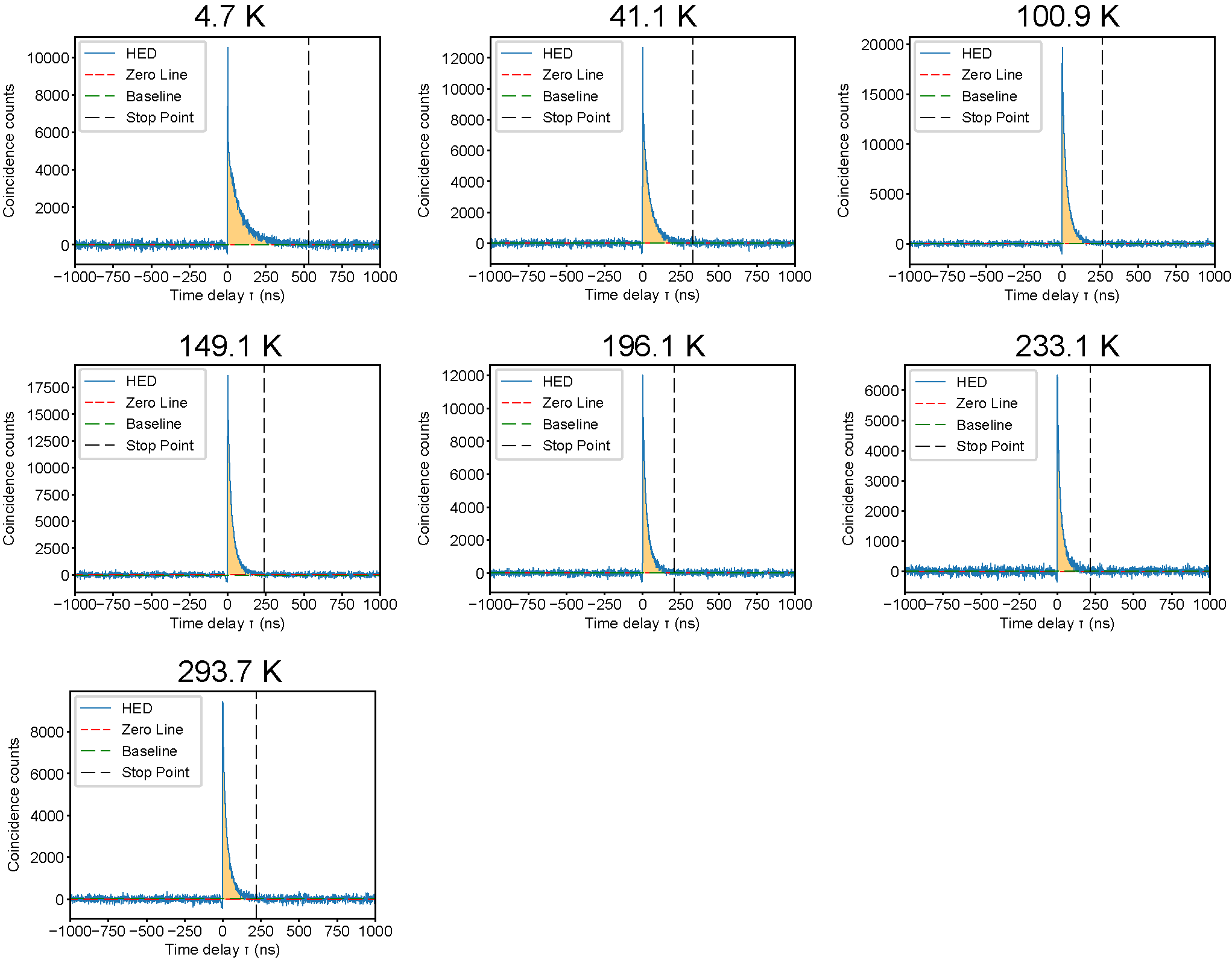}
\caption{Accidentally corrected coincidence counts of the HED data for Sample 1. The red line indicates zero, and the green line represents the baseline. The shaded area corresponds to the summation of the bunching peak, while the black line marks the end of the summation.}
\label{figSI_s1_subacc}
\end{figure}

\begin{figure}[H]
\centering
\includegraphics[width=16cm]{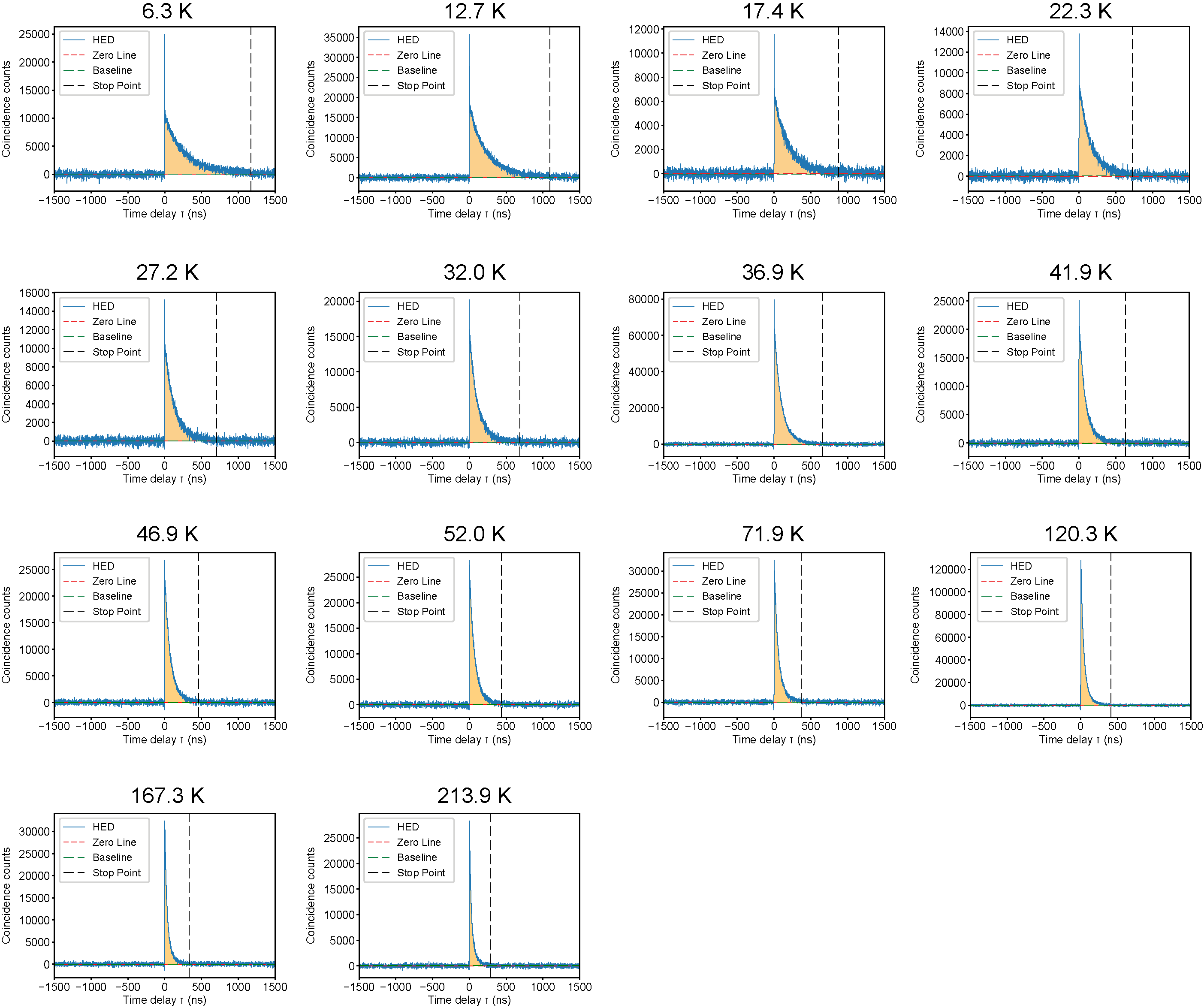}
\caption{Accidentally corrected coincidence counts of the HED data for Sample 2. The red line indicates zero, and the green line represents the baseline. The shaded area corresponds to the summation of the bunching peak, while the black line marks the end of the summation.}
\label{figSI_s2_subacc}
\end{figure}

Here, we report the coincidence counts between the heralding and emission photons for Sample 1 and Sample 2 as shown in Fig. \ref{figSI_s1_subacc} and Fig. \ref{figSI_s2_subacc}, respectively. The coincidence counts were extracted from the bunching peak of the raw coincidence histogram, $G^{(2)}(\tau)$, after subtracting accidental coincidences. To determine the coincidence region, we defined a moving average window of 10 points, and the baseline $B$ was estimated as the median of the first and last 50 points of the accidental-corrected counts. Starting from the maximum peak position, we iteratively moved forward in time, calculating the moving average over the defined window at each step. The summation of counts continued until the moving average dropped below the baseline, at which point the stopping index was recorded. This procedure ensures that only counts associated with the main bunching peak—those significantly above the background—are included in the reported coincidence values. The results are shown in Table \ref{s1_coin}, and \ref{s2_coin}. 

\begin{table}[h]
    \centering
    \small 
    \begin{threeparttable} 
    \caption{Coincidence counts for Sample 1.}
    \label{s1_coin}
        \begin{tabular}{c c c c}
            \hline
            Temperature (K) & Coincidences (counts) & Collection time (seconds) & Coincidence rate ($cps$)\\
            \hline
            4.7 & 444501.69 & 54000 & 8.23 \\ 
            41.1& 413424.97 & 64800 & 6.38 \\
            100.9 & 560339.76 & 64800 & 8.65\\
            149.1 & 490492.96 & 64800 & 7.57\\
            196.1 & 306575.17 & 53999 & 5.68\\
            233.1 & 159341.18 & 54000 & 2.95\\
            293.7 & 262450.01 & 52539 & 5.00\\
            \hline
        \end{tabular}
    \end{threeparttable}
\end{table}

\begin{table}[h]
    \centering
    \small 
    \begin{threeparttable} 
        \caption{Coincidence counts for Sample 2.}
        \label{s2_coin}
        \begin{tabular}{c c c c}
            \hline
            Temperature (K) & Coincidences (counts) & Collection time (seconds) & Coincidence rate ($cps$)\\
            \hline
            6.3 & 2936744.11 & 14399 & 203.95 \\ 
            12.7 &4240921.85 & 21600 & 196.34 \\
            17.4 & 1290697.47 & 7200 & 179.26\\
            22.3 & 1424886.98 & 7199 & 197.93\\
            27.2 & 1391828.09 & 7199 & 193.34\\
            32.0 & 1782515.28 & 8999 & 198.08\\
            36.9 & 6533190.49 & 23399 & 279.21\\
            41.9 & 1928697.66 & 7199 & 267.91\\
            46.9 & 1940265.14 & 7199 & 269.52\\
            52.0 & 1997718.89 & 7199 & 277.50\\
            71.9 & 1939763.46 & 7199 & 269.45\\
            120.3 & 6142470.33 & 23399 & 262.51\\
            167.3 & 1289089.44 & 7199 & 179.07\\
            213.9 & 970471.83 & 7199 & 134.81\\
            \hline
        \end{tabular}
    \end{threeparttable}
\end{table}

\subsection{Downonversion efficiency calculation}
The downconversion efficiency is defined as the ratio of emitted photons to excited photons. We calculated this efficiency for Samples 1 and 2, and the results are summarized in Tables~\ref{s1_eff} and \ref{s2_eff}. The efficiency ranges from approximately $10^{-5}$ to $10^{-3}$. The relatively low conversion efficiency observed for Sample 1 is primarily attributed to limited coupling efficiency into the single-mode fiber.

\begin{table}[h]
    \centering
    \footnotesize 
    \setlength{\tabcolsep}{4pt} 
    \renewcommand{\arraystretch}{1.1} 
    \begin{threeparttable}
        \caption{Emission photon conversion efficiency of Sample 1.}
        \label{s1_eff}
        \begin{tabular}{c c c c c}
            \hline
            Temperature (K) & Heralding count rate (Mcps) & Excitation count rate (Mcps) & Emission count rate (cps) & Efficiency \\
            \hline
            4.7   & 3.41& 1.54 & 89.07 & $5.80 \times 10^{-5}$ \\ 
            41.1  & 3.32 & 1.49& 75.15 & $5.03 \times 10^{-5}$ \\
            100.9 & 3.37 & 1.52& 104.01 & $6.86 \times 10^{-5}$ \\
            149.1   & 3.30& 1.49& 94.07 & $6.33 \times 10^{-5}$ \\
            196.1   & 3.12& 1.40& 75.40 & $5.38 \times 10^{-5}$ \\
            233.1   & 3.09 & 1.39& 43.22 & $5.42 \times 10^{-5}$ \\
            293.7   & 2.75& 1.24& 69.81 & $5.64 \times 10^{-5}$ \\
            \hline
        \end{tabular}
    \end{threeparttable}
\end{table}


\begin{table}[h]
    \centering
    \footnotesize 
    \begin{threeparttable} 
        \caption{Emission photon conversion efficiency of Sample 2.}
        \label{s2_eff}
        \begin{tabular}{c c c c c }
            \hline
            Temperature (K) & Heralding count rate (Mcps) & Excitation count rate (Mcps) & Emission count rate (kcps) & Efficiency \\
            \hline
            6.3   & $2.534$ & $2.281$ & 3.773 & $1.65\times 10^{-3}$ \\ 
            12.7  & $2.443$ & $2.199$ & 3.661 & $1.67\times 10^{-3}$ \\
            17.4   & $2.461$ & $2.215$ & 3.479 & $1.57\times 10^{-3}$\\
            22.3   & $2.513$ & $2.262$ & 3.630  & $1.61\times 10^{-3}$ \\
            27.2   & $2.612$ & $2.351$ & 3.637 & $1.55\times 10^{-3}$ \\
            32.0   & $2.675$ & $2.408$ & 3.646  & $1.51\times 10^{-3}$ \\
            36.9   & $2.686$ & $2.417$ & 4.351 & $1.80\times 10^{-3}$ \\
            41.9   & $2.700$ & $2.430$ & 4.232 & $1.74\times 10^{-3}$ \\
            46.9   & $2.704$ & $2.434$ & 4.276 & $1.76\times 10^{-3}$ \\
            52.0   & $2.725$ & $2.453$ & 4.328 & $1.76\times 10^{-3}$ \\
            71.9   & $2.735$ & $2.462$ & 4.301 & $1.75\times 10^{-3}$ \\
            120.3   & $2.747$ & $2.472$ & 4.205 & $1.70\times 10^{-3}$ \\
            167.3   & $2.721$ & $2.449$ & 3.491 & $1.43\times 10^{-3}$ \\
            213.9   & $2.762$ & $2.486$ & 3.111 & $1.25\times 10^{-3}$ \\
            \hline
        \end{tabular}
    \end{threeparttable}
\end{table}

\subsection{Comparison between HED and TCSPC}
To assess whether the temporal decay profiles obtained from the HED measurement and the TCSPC technique exhibit consistent behavior, we compared their respective photon correlation and lifetime traces using data from Sample 1. A direct comparison between these two methods is not straightforward, as the photon correlation function derived from HED data typically contains contributions from uncorrelated background photons, whereas TCSPC employs a start–stop detection scheme that results in a different baseline behavior. To enable a meaningful comparison—particularly on a logarithmic scale—we analyzed the coincidence counts after correcting for accidental coincidences, as shown in the previous section.

\begin{figure}[h]
\centering\includegraphics[width=10cm]{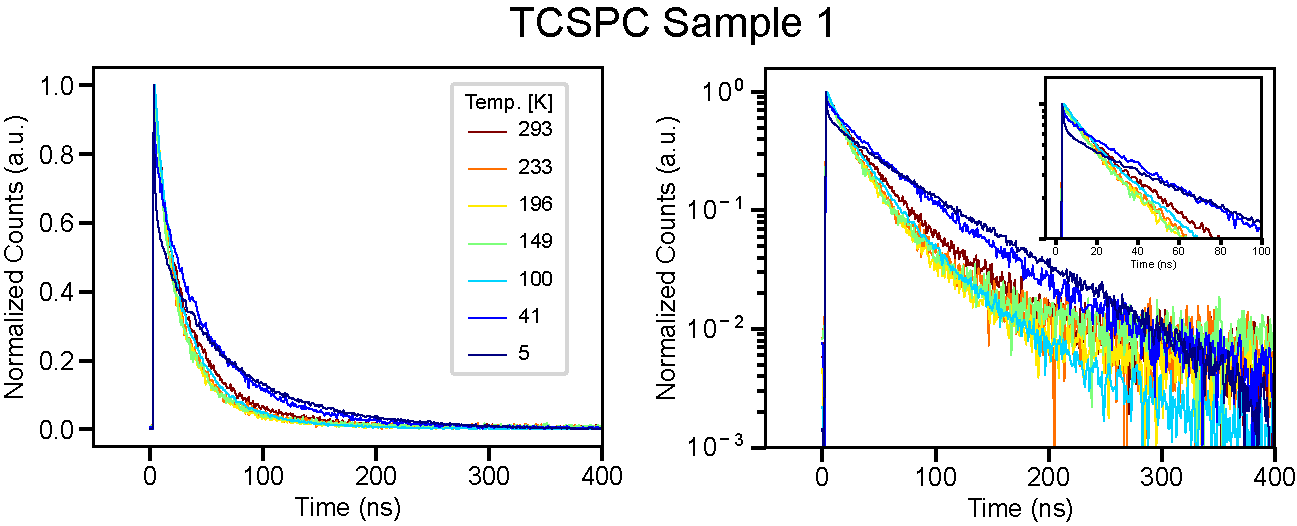}
\caption{TCSPC data of Sample 1. The left panel shows the data on a linear scale, while the right panel shows the data on a logarithmic scale, with a zoomed-in view displayed in the inset.}
\label{figSI_s1_TCSPC}
\end{figure}

\begin{figure}[h]
\centering\includegraphics[width=15cm]{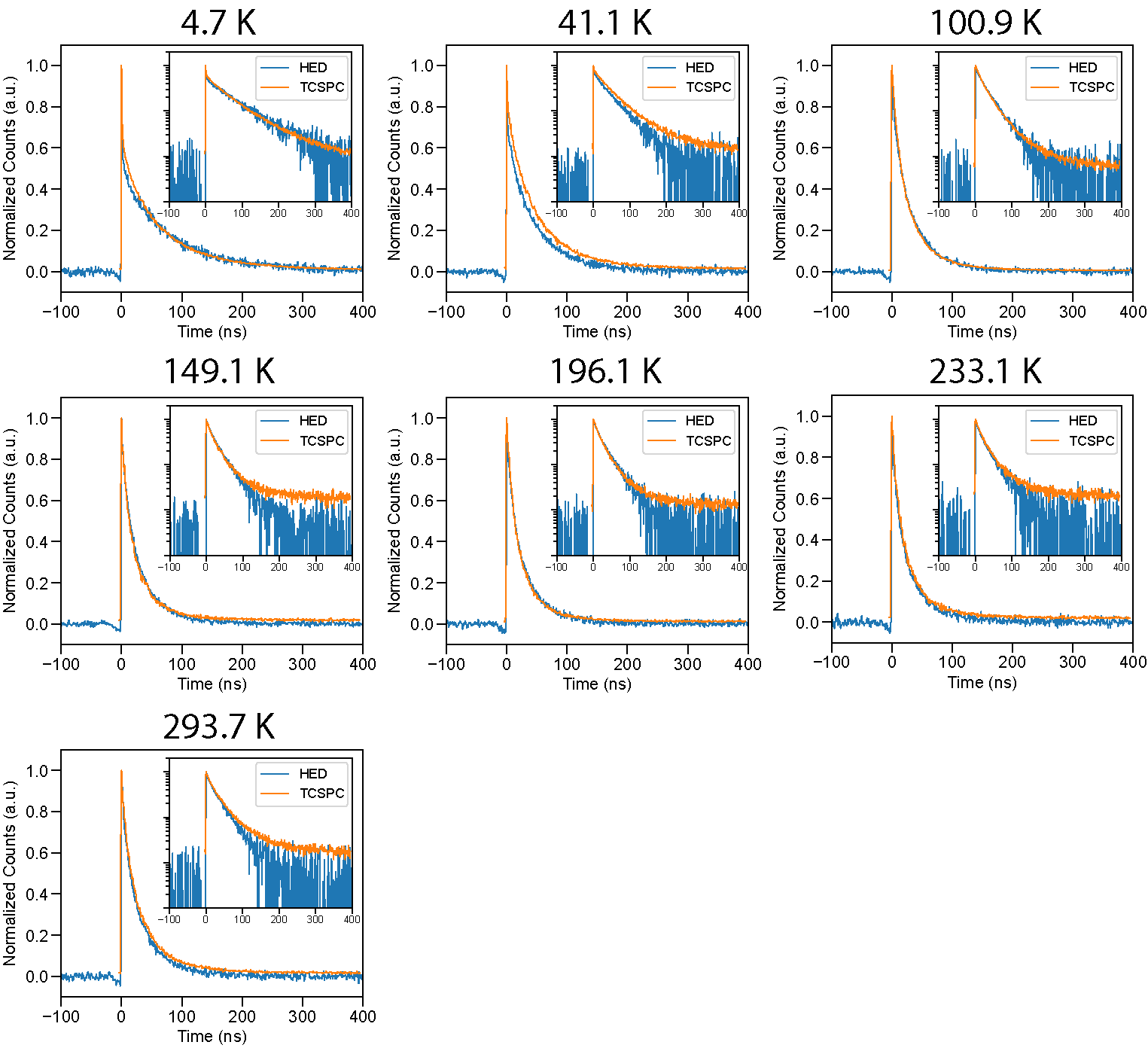}
\caption{Overlay of the HED and TCSPC data for Sample 1 under different temperatures.}
\label{figSI_s1_HEDvsTCSPC}
\end{figure}

TCSPC measurements were performed at the same sample position using a 470~nm pulsed laser (Fig. \ref{figSI_s1_TCSPC}). For direct comparison, the normalized HED and TCSPC traces were overlaid after temporal alignment. This comparison revealed highly consistent decay dynamics between the two methods (Fig. \ref{figSI_s1_HEDvsTCSPC}), demonstrating that both techniques capture similar underlying emission kinetics.

To quantify the similarity between two datasets \(A\) and \(B\), we employed three common metrics: the Pearson correlation coefficient (PCC), the mean squared error (MSE), and the normalized cross-correlation (NCC). The PCC measures the degree of linear correlation between two variables and is defined as
\begin{equation}
\mathrm{PCC} = \frac{\sum_i (A_i - \bar{A})(B_i - \bar{B})}{\sqrt{\sum_i (A_i - \bar{A})^2 \sum_i (B_i - \bar{B})^2}},
\end{equation}
where \(\bar{A}\) and \(\bar{B}\) are the mean values of \(A\) and \(B\), respectively. The PCC ranges from \(-1\) to \(1\), with \(1\) indicating perfect positive linear correlation.

The MSE quantifies the average squared difference between corresponding elements of the two datasets and is given by
\begin{equation}
\mathrm{MSE} = \frac{1}{N} \sum_i (A_i - B_i)^2,
\end{equation}
where \(N\) is the total number of data points. A smaller MSE value indicates greater similarity between \(A\) and \(B\).

The NCC evaluates the similarity between two normalized datasets and is defined as
\begin{equation}
\mathrm{NCC} = \frac{\sum_i (A_i - \bar{A})(B_i - \bar{B})}{N \, \sigma_A \sigma_B},
\end{equation}
where \(\sigma_A\) and \(\sigma_B\) denote the standard deviations of \(A\) and \(B\), respectively. The NCC value also ranges from \(-1\) to \(1\), with higher values corresponding to stronger similarity. In practice, PCC and NCC provide comparable measures of correlation-based similarity, whereas MSE emphasizes absolute intensity differences.

When each dataset is normalized by its maximum value (i.e., \(A' = A / A_{\max}\) and \(B' = B / B_{\max}\)), the three similarity metrics respond differently. Because PCC and NCC depend only on relative variations after mean subtraction, they are invariant under linear scaling; thus, dividing by the maximum value does not alter their results. In contrast, MSE is sensitive to absolute magnitudes—scaling both datasets by their respective maxima changes the overall amplitude and typically reduces the MSE value. Therefore, while PCC and NCC remain unaffected by such normalization, MSE decreases proportionally with the applied scaling.

We applied these three similarity metrics to evaluate the agreement between the HED and TCSPC decay traces across different temperatures, as summarized in Table~\ref{s1_HEDvsTCPSC}. The consistently high PCC and NCC values (\(>0.95\)) and the low MSE values confirm the excellent correspondence between the two methods.

\begin{table}[h]
    \centering
    \caption{Similarity metrics between HED and TCSPC decay traces at various temperatures.}
    \label{s1_HEDvsTCPSC}
    \small
        \begin{tabular}{c c c c}
            \hline
            Temperature (K) & PCC & MSE & NCC \\
            \hline
            4.7   & 0.991 & $4.98 \times 10^{-4}$ & 0.991 \\ 
            41.1  & 0.995 & $1.63 \times 10^{-3}$ & 0.995 \\
            100.9 & 0.962 & $2.12 \times 10^{-3}$ & 0.962 \\
            149.1 & 0.960 & $1.98 \times 10^{-3}$ & 0.960 \\
            196.1 & 0.957 & $2.10 \times 10^{-3}$ & 0.957 \\
            233.1 & 0.957 & $2.52 \times 10^{-3}$ & 0.957 \\
            293.7 & 0.967 & $2.22 \times 10^{-3}$ & 0.967 \\
            \hline
        \end{tabular}
\end{table}

\subsection{Lifetime extraction from HED and TCSPC}
The similarity analysis described above is sometimes indirect and may not fully capture differences in decay dynamics. To enable a more quantitative comparison, we fitted the HED and TCSPC data of Sample 1 to extract average lifetimes. All decay traces were fitted using bi- and tri-exponential models over a consistent time range of approximately 400 ns for both HED and TCSPC data.

Generally, the average lifetime of a multiexponential photoluminescence decay can be defined in several ways depending on the weighting scheme. For a decay described by
\begin{equation}
I(t) = \sum_i A_i e^{-t / \tau_i},
\end{equation}
where $A_i$ and $\tau_i$ are the pre-exponential amplitude and decay constant of the $i$-th component, respectively, the \textit{intensity-weighted} (or population-weighted) average lifetime is given by
\begin{equation}
\langle \tau \rangle_{\mathrm{int}} = \frac{\sum_i A_i \tau_i^2}{\sum_i A_i \tau_i}.
\end{equation}
This definition originates from the first moment of the decay curve, $\langle \tau \rangle = \int_0^{\infty} t I(t) \, dt \, / \, \int_0^{\infty} I(t) \, dt$, and physically represents the mean emission time per emitted photon. Each exponential term contributes an integrated photon intensity proportional to $A_i \tau_i$; thus, weighting by $A_i \tau_i$ properly accounts for the total photon flux of each component. The intensity-weighted lifetime provides a more physically meaningful measure of the effective emission lifetime, as it reflects the photon-emission–weighted average decay time.

For our data, the biexponential model adequately reproduces the HED results across the entire temperature range. In contrast, for TCSPC data acquired below 100.9~K, this model fails to fully capture the decay dynamics—particularly at 4.7, 41.1, and 100.9~K. In these cases, a triexponential model provides a significantly improved fit, suggesting the emergence of additional relaxation pathways at low temperatures. The average lifetimes obtained from both bi- and triexponential fits are summarized in Table~\ref{s1_HEDvsTCSPC_lifetime}. Overall, the HED and TCSPC results show good agreement in the average lifetime values across most temperatures, except at 4.7~K.

\begin{table}[h]
  \centering
  \caption{Average lifetimes extracted from bi- and triexponential fits for HED and TCSPC measurements of Sample~1.}
  \label{s1_HEDvsTCSPC_lifetime}
  \begin{threeparttable}
    \begin{tabular}{ccccc}
      \hline
      \multirow{3}{*}{Temperature (K)} & \multicolumn{4}{c}{Average lifetime (ns)} \\
                                       & \multicolumn{2}{c}{Biexp fitting} & \multicolumn{2}{c}{Triexp fitting} \\
                                       & HED & TCSPC & HED & TCSPC \\
      \hline
      4.7   & \ensuremath{74.05 \pm 1.03} & \ensuremath{61.73 \pm 0.36} & \ensuremath{77.64 \pm 1.51} & \ensuremath{68.60 \pm 0.77} \\
      41.4  & \ensuremath{44.51 \pm 0.42} & \ensuremath{48.52 \pm 0.30} & \ensuremath{48.35 \pm 1.53} & \ensuremath{52.72 \pm 1.21} \\
      100.9 & \ensuremath{33.83\pm 0.47} & \ensuremath{32.98 \pm 0.44} & \ensuremath{36.00 \pm 1.78} & \ensuremath{38.91 \pm 10.47} \\
      149.1 & \ensuremath{31.98 \pm 0.76} & \ensuremath{29.84 \pm 0.95} & \ensuremath{33.00 \pm 1.31} & \ensuremath{30.55 \pm -} \\
      196.1 & \ensuremath{31.65 \pm 0.97} & \ensuremath{29.07 \pm 0.84} & \ensuremath{32.78 \pm 1.89} & \ensuremath{33.30 \pm 18.85} \\
      233.1 & \ensuremath{32.10 \pm 1.50} & \ensuremath{30.69 \pm 1.03} & \ensuremath{34.40 \pm 3.31} & \ensuremath{35.18 \pm 8.55} \\
      293.7 & \ensuremath{34.13 \pm 0.85} & \ensuremath{36.60 \pm 0.74} & \ensuremath{36.77 \pm 3.97} & \ensuremath{37.52 \pm 1.05} \\
      \hline
    \end{tabular}
    \begin{tablenotes}[flushleft]
      \footnotesize
      \item Note: Uncertainties represent one standard deviation from the fit. A ``-'' indicates an unphysical or undefined standard deviation.
    \end{tablenotes}
  \end{threeparttable}
\end{table}

From both qualitative and quantitative analyses, subtle differences remain in the decay profiles, particularly in the long-time regime. Moreover, noticeable discrepancies in average lifetime values persist at low temperatures. These differences may arise from several factors: (1) the thermal photon distribution in the HED sample arm, which differs from the Poissonian statistics of pulsed excitation; (2) the 470~nm pulsed excitation in TCSPC, which exceeds the bandgap energy and can generate a higher density of free carriers; and (3) slight spatial mismatches between the regions probed by HED and TCSPC, resulting in variations in local optical properties.

For Sample 2, we performed single-, bi-, and tri-exponential fittings to extract the lifetimes and each component, respectively. Similar to Sample 1, as the temperature increases, the fastest decay component gradually disappears, resulting in a decrease in the total average lifetime. For temperatures $\le 52.0$~K, we employed a biexponential model, which provides a more reliable estimate of the average lifetime, as a single-exponential fit cannot capture the fastest decay component and therefore underestimates the lifetime. Although the tri-exponential model can mathematically resolve all components, strong correlations between parameters make it unsuitable for analyzing individual decay components. Consequently, we relied on the biexponential fit for the lower-temperature data. At higher temperatures, all three models yield similar lifetimes; to maintain consistency with the lower-temperature analysis and account for the disappearance of the fastest decay component, we used a single-exponential fit for the remaining decay. Overall, the biexponential model was applied for lower-temperature data, and the single-exponential model for higher-temperature data in the subsequent analysis.

\begin{table}[h]
  \centering
  \caption{Average lifetimes extracted from single-, bi-, and triexponential fits for Sample~2 HED data.}
  \label{s2_HED_lifetime}
  \begin{threeparttable}
    \begin{tabular}{cccc}
      \hline
      \multirow{2}{*}{Temperature (K)} & \multicolumn{3}{c}{Average lifetime (ns)} \\
                                       & Single exp fitting & Biexp fitting & Triexp fitting \\
      \hline
      6.3   & \ensuremath{245.74 \pm 3.74} & \ensuremath{257.81 \pm 2.78} & \ensuremath{261.93 \pm 3.23} \\
      12.7  & \ensuremath{218.15 \pm 2.30} & \ensuremath{225.96 \pm 1.61} & \ensuremath{242.91 \pm 26.18} \\
      17.4 & \ensuremath{183.85 \pm 2.18} & \ensuremath{188.92 \pm 2.02} & \ensuremath{198.81 \pm 43.05} \\
      22.3& \ensuremath{151.86 \pm 1.37} & \ensuremath{155.19 \pm 1.27} & \ensuremath{160.57 \pm 47.96} \\
      27.2 & \ensuremath{128.13 \pm 1.07} & \ensuremath{130.57 \pm 1.00} & \ensuremath{143.51 \pm 78.07} \\
      32.0& \ensuremath{107.11 \pm 0.67} & \ensuremath{108.37 \pm 0.65} & \ensuremath{114.69 \pm 24.10} \\
      36.9 & \ensuremath{94.29 \pm 0.31} & \ensuremath{95.19 \pm 0.27} & \ensuremath{112.96 \pm 15.09} \\
      41.9 & \ensuremath{84.61 \pm 0.42} & \ensuremath{85.28 \pm 0.66} & \ensuremath{95.49 \pm -} \\
      46.9 & \ensuremath{77.57 \pm 0.35} & \ensuremath{78.04 \pm 0.37} & \ensuremath{89.90 \pm -} \\
      52.0 & \ensuremath{71.79 \pm 0.29} & \ensuremath{72.10 \pm 0.30} & \ensuremath{82.96 \pm 12.34} \\
      71.9 & \ensuremath{58.05 \pm 0.23} & \ensuremath{68.37 \pm 7.71} & \ensuremath{65.52 \pm 6.78} \\
      120.3 & \ensuremath{44.23 \pm 0.13} & \ensuremath{54.96 \pm 2.04} & \ensuremath{54.13 \pm 1.93} \\
      167.3 & \ensuremath{37.31 \pm 0.19} & \ensuremath{46.73 \pm 2.38} & \ensuremath{76.67 \pm 179.13} \\
      213.9 & \ensuremath{32.48 \pm 0.21} & \ensuremath{43.27 \pm 1.83} & \ensuremath{44.55 \pm 11.36} \\
      \hline
    \end{tabular}
    \begin{tablenotes}[flushleft]
      \footnotesize
      \item Note: Uncertainties represent one standard deviation from the fit. A ``-'' indicates an unphysical or undefined standard deviation.
    \end{tablenotes}
  \end{threeparttable}
\end{table}

\subsection{Decay compoentns analysis}
\subsubsection{Fastest decay component ratio versus temperature}

\begin{figure}[h]
\centering
\includegraphics[width=10cm]{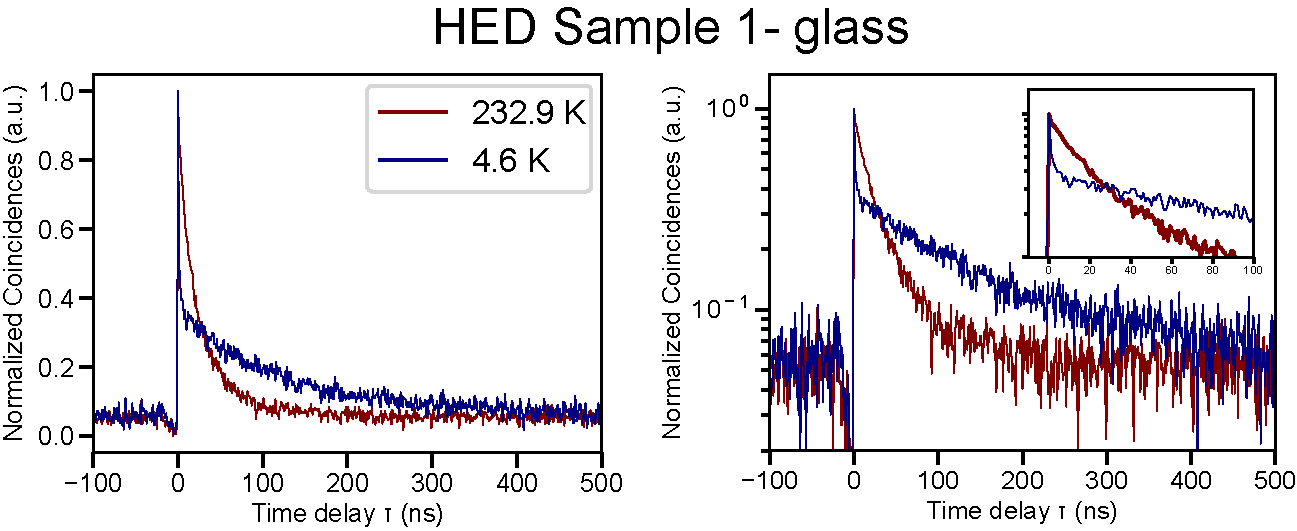}
\caption{HED data of Sample 1 on a glass substrate. The left panel shows the data on a linear scale, while the right panel presents the data on a logarithmic scale, with a zoomed-in view displayed in the inset.}
\label{figSI_s1_nogold}
\end{figure}

We observed an initial fast decay component in our data. We have ruled out the possibility that this decay arises from nonradiative processes due to the gold mirror contacting the QDs, as a similar measurement performed on a control sample with a glass substrate instead of a gold mirror shows the same behavior (Fig.~\ref{figSI_s1_nogold}). 

\begin{figure}[h]
\centering\includegraphics[width=8cm]{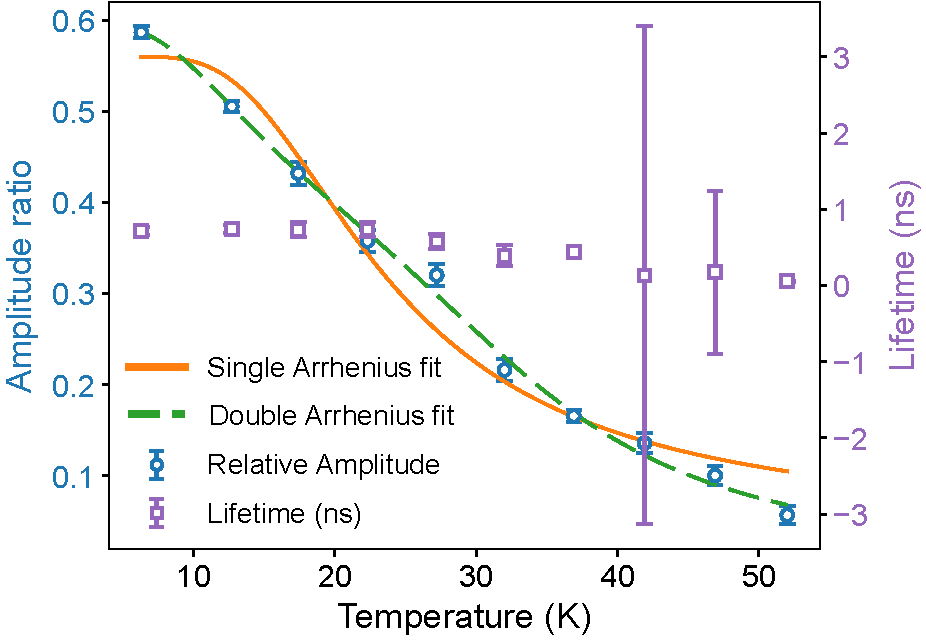}
\caption{Amplitude ratio and lifetime plot of Sample 2 from HED data. The lifetime remains approximately constant, while the amplitude ratio varies significantly. The data are fitted using single- and double-Arrhenius models, with the double-Arrhenius fit providing a better representation.}
\label{figSI_s2_HED_fast_decay}
\end{figure}

Therefore, this fast decay may originate from other nonradiative recombination channels or from an additional relaxation process, such as the transition from the bright to the dark state. This component is progressively quenched as the temperature increases. Its lifetime remains nearly constant until the component becomes undetectable, while its relative amplitude varies with temperature, as shown in Fig.~\ref{figSI_s2_HED_fast_decay}. To quantify this behavior, we fitted the relative amplitude as a function of temperature up to 52.0~K, the highest temperature at which the initial fast component is still observable. To model this decay, we apply an Arrhenius-type model, which accounts for thermally activated quenching processes. The temperature dependence of the relative amplitude $A(T)$ is described by
\begin{equation}
A(T) = \frac{A_0}{1 + \sum_i C_i \, e^{-E_i/(k_B T)}},
\end{equation}
where $A(T)$ is the relative amplitude of the fast decay component at temperature $T$, $A_0$ is the amplitude at the lowest temperature, $C_i$ are scaling factors associated with the quenching processes, $E_i$ are the activation energies of the $i$-th process, and $k_B$ is the Boltzmann constant.

We found that this double-Arrhenius model provides a better description of the temperature dependence than a single-process model, as shown in Fig.~\ref{figSI_s2_HED_fast_decay}.

From the fitting, we obtained $A_0 = 0.593$, $C_1 = 225.3$, $E_1 = 15.99$~meV, $C_2 = 2.62$, and $E_2 = 2.96$~meV. The parameter $A_0$ represents the relative amplitude of the fast decay component at the lowest temperature, reflecting its intrinsic contribution. The two terms in the denominator correspond to two thermally activated quenching processes: the first process, with a large scaling factor $C_1$ and higher activation energy $E_1$, dominates the quenching at intermediate temperatures, while the second process, with a smaller scaling factor $C_2$ and lower activation energy $E_2$, contributes to quenching even at very low temperatures. This indicates that the fast decay component is influenced by multiple trap-mediated pathways, with one process requiring more thermal energy to activate than the other, consistent with the observed temperature dependence in Fig.~\ref{figSI_s2_HED_fast_decay}.

\subsubsection{Remaining lifetime versus temperature}

\begin{figure}[H]
\centering\includegraphics[width=10cm]{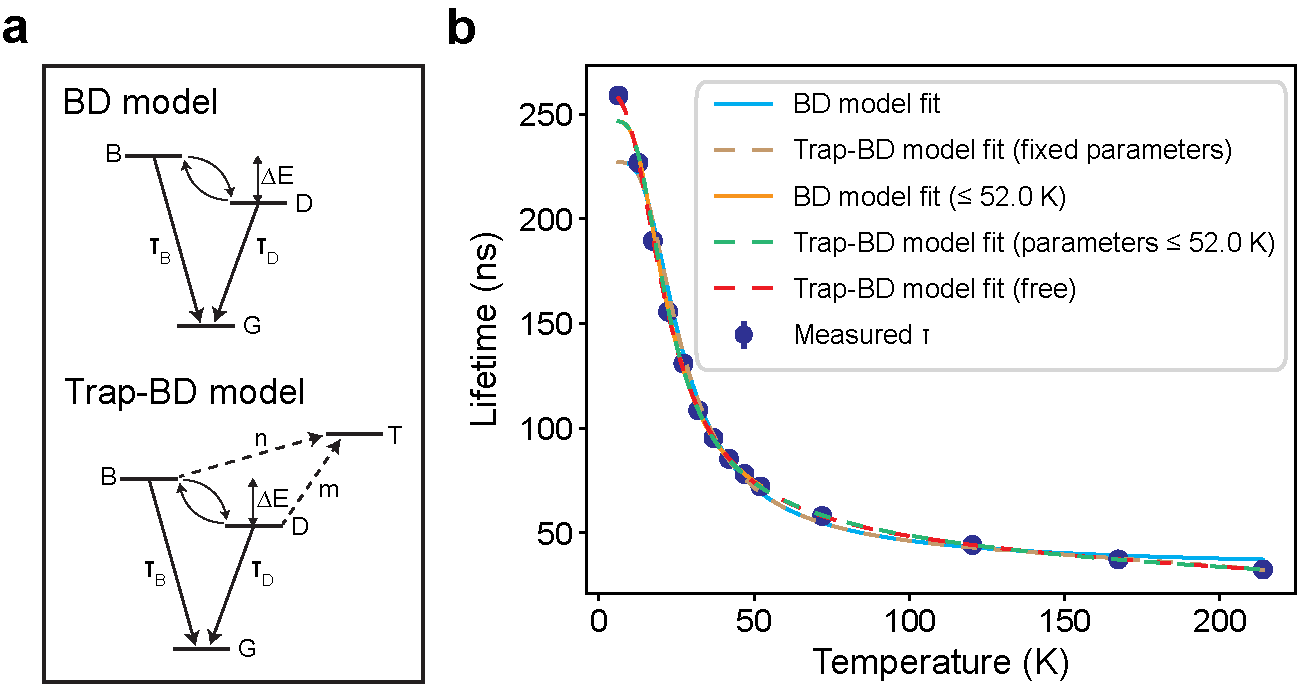}
\caption{Model for the remaining decay fitting. 
(a) Schematic of the bright–dark (BD) model and the trap–BD state model. $G$ is the ground state, $B$ is the bright state, $D$ is the dark state, and $T$ is the trap state. $\tau_B$ and $\tau_D$ are the lifetimes of the bright and dark states, respectively. $\Delta E$ is the energy splitting between the bright and dark states. $m$ and $n$ refer to the number of photons required for carrier escape from the dark and bright states. 
(b) Comparison of different model fits for lifetime as a function of temperature.}
\label{figSI_s2_HED_BD}
\end{figure}

To model the temperature dependence of the remaining lifetimes, we first applied the bright–dark (BD) state model. In this framework, the average lifetime is determined by the population distribution between the bright and dark states, as illustrated in Fig.~\ref{figSI_s2_HED_BD}. The temperature-dependent average lifetime is expressed as

\begin{equation}
\tau_\mathrm{avg}^{-1} =
\frac{e^{\Delta E / (k_B T)}}{1 + e^{\Delta E / (k_B T)}} \tau_D^{-1} +
\frac{1}{1 + e^{\Delta E / (k_B T)}} \tau_B^{-1},
\label{eq:BD_model}
\end{equation}

where $\tau_B$ and $\tau_D$ are the lifetimes of the bright and dark states, respectively, and $\Delta E$ is the energy splitting between them. As shown by the red curve in Fig.~\ref{figSI_s2_HED_BD}, the BD model does not fully capture the behavior at the lowest and highest temperatures, suggesting the presence of additional decay mechanisms. Even when restricting the fit to temperatures $\leq 52.0$~K, the model still deviates slightly at the lowest temperatures.

Since defects are often more prevalent in core–shell structures and can act as trap states, we extended the BD model to include a trap-state contribution, following Refs.~\cite{murphy2016temperature, gaponenko2010temperature}. The resulting trap–BD model is expressed as

\begin{equation}
\tau_\mathrm{avg}^{-1} = \frac{a \tau_B^{-1} + \tau_D^{-1} + k_0 \left[ a (b-1)^{-n} + (b-1)^{-m} \right]}{1 + a},
\end{equation}

with
\begin{align}
a = e^{-\Delta E / (k_B T)}, \quad
b = e^{E_\mathrm{ph} / (k_B T)},
\end{align}

where $k_0$ is the trap-assisted decay rate, $E_\mathrm{ph}$ is the relevant phonon energy, and $n$ and $m$ are phonon interaction exponents ($m>n$). Following the procedure in Refs.~\cite{murphy2016temperature, gaponenko2010temperature}, we initially fixed $\tau_B$, $\tau_D$, and $\Delta E$ to the values obtained from the BD model fitted across all temperatures, treating $k_0$, $m$, and $n$ as free parameters. In this regime, the fit reproduces the BD model at low temperatures while improving agreement at higher temperatures. We also tested using BD parameters obtained from fitting $T \leq 52.0$~K, which gave similar trends. A subsequent fully unconstrained fit, allowing $\tau_B$, $\tau_D$, and $\Delta E$ to vary, yielded excellent agreement across the entire temperature range; however, some parameter values were unphysical (e.g., $m$ was not larger than $n$ as required by the model). Nonetheless, the trap–BD model provides a more accurate description of the temperature-dependent lifetime across the full temperature range.

All fitting parameters are listed in Table~\ref{SI_BD}. Because $m$ and $n$ are too close in the trap–BD fit using fixed BD parameters, and the fully unconstrained trap–BD fit is unreliable, the trap–BD fit with BD parameters fixed for $T \leq 52.0$~K provides more reliable and physically meaningful values.

\begin{table}[h]
    \centering
    \scriptsize
    \caption{Fitting parameters obtained from different models applied to the HED data of Sample 2.}
    \label{SI_BD}
    \begin{threeparttable}
        \begin{tabular}{c c c c c c c c}
            \hline
            Model & $\tau_B$ (ns) & $\tau_D$ (ns) & $\Delta E$ (meV) &$E_\mathrm{ph}$ (meV)& $1/k_0$ (ns)&$m$&$n$ \\
            \hline
            BD   & 16.93 & 227.02 & 6.69 & - & - & - & -\\
            Trap-BD (fixed BD parameters) & 16.93 & 227.02 & 6.69 & 153.71 & 1.02 & 0.68 & 0.63 \\
            BD ($\leq 52.0$~K)  & 21.36 & 246.62 & 5.49 & - & - & - & -\\
            Trap-BD (fixed BD ($\leq 52.0$~K) parameters) & 21.36 & 246.62 & 5.49 & 23.84 & 1.03 & 1.03 & 0.79 \\
            Trap-BD &  134.56& 260.03& 2.49&28.70&19.39&0.21&1.30 \\
            \hline
        \end{tabular}
    \end{threeparttable}
\end{table}

Overall, it appears that multiple defects and phonon-mediated channels, each with distinct activation energies, contribute to the complex decay observed in HED measurements. The pronounced Stokes shift further suggests the presence of numerous relaxation pathways.


\section*{References}

\printbibliography[heading=none]
\end{refsection}

\end{document}